\documentclass{ws-ijmpc}

\begin{document}

\markboth{Rafael M. Brum, Nuno Crokidakis}
{Dynamics of tax evasion through an epidemic-like model}

\catchline{}{}{}{}{}

\title{Dynamics of tax evasion through an epidemic-like model}

\author{Rafael M. Brum, Nuno Crokidakis $^{*}$}

\address{
Instituto de F\'{\i}sica, \hspace{1mm} Universidade Federal Fluminense \\
Av. Litor\^anea s/n, \hspace{1mm} 24210-340 \hspace{1mm} Niter\'oi - Rio de Janeiro, \hspace{1mm} Brazil \\ 
$^{*}$ Corresponding author: nuno@if.uff.br}

\maketitle

\begin{history}
\received{Day Month Year}
\revised{Day Month Year}
\end{history}

\begin{abstract}
\noindent
In this work we study a model of tax evasion. We considered a fixed population divided in three compartments, namely honest tax payers, tax evaders and a third class between the mentioned two, which we call \textit{susceptibles} to become evaders. The transitions among those compartments are ruled by probabilities, similarly to a model of epidemic spreading. These probabilities model social interactions among the individuals, as well as the government's fiscalization. We simulate the model on fully-connected graphs, as well as on scale-free and random complex networks. For the fully-connected and random graph cases we observe that the emergence of tax evaders in the population is associated with an active-absorbing nonequilibrium phase transition, that is absent in scale-free networks.


\end{abstract}

\ccode{PACS Nos.: 05.10.-a, 05.70.Jk, 87.23.Ge, 89.75.Fb}

\section{Introduction}

\qquad Socio-economic problems have been the targets of several studies in last years \cite{econ_book,pmco_book}. Those interdisciplinary topics are usually treated by means of computer simulations of agent-based models, which allow us to understand the emergence of collective phenomena in those systems. Among the studied problems, one of great interest is tax evasion dynamics, which is interesting from the practical point of view because tax evasion remains to be a major predicament facing governments \cite{bloom,prinz,andreoni}. Economists studied models of tax evasion during several years \cite{gachter,frey,follmer,slemrod,davis,salmina,wenzel,hood}, and more recently physicists also became interested in the subject \cite{zaklan,lima1,lima2,lima3,llacer,seibold,bertotti,meu_econo} (for recent reviews, see \cite{bloom,prinz}). Experimental evidence provided by Gachter suggests that tax payers tend to condition their decision regarding whether to pay taxes or not on the tax evasion decision of the members of their group \cite{gachter}. In addition, Frey and Torgler also provide empirical evidence on the relevance of conditional cooperation for tax morale \cite{frey}. Taking those discussions into account, Zaklan \textit{et al.} recently proposed a model that has been attracted attention \cite{zaklan}. In the so-called Zaklan model, the dynamics of tax payers and tax evaders is analyzed by means of the two-dimensional Ising model at a given temperature $T$. Each agent $i$ in the artificial population may be in one of two possible states, namely $s_{i}=+1$ (honest) or $s_{i}=-1$ (cheater or tax evader). A transition $s_{i} \to -s_{i}$ (or a spin flip) is controlled by the ``social temperature'' $T$ and also depends on the nearest neighbors' states of the agent (or spin) at site $i$. Thus, for low temperatures few spin flips occur and for high temperatures many spin flips occur. In other words, tax evaders have the greatest influence to turn honest citizens into tax evaders if they constitute a majority in the respective neighborhood. In addition, some punishment rules are applied: there is a probability $p_{a}$ of an audit each agent is subject to in every period and a length of time $k$ detected tax evaders remain honest \cite{zaklan}. In another work, the dynamics of the model was also controlled by another two-state model, namely the majority-vote model with noise \cite{maj_vot}, where the noise $q$ plays the role of the temperature. In this case, similar results were found \cite{lima3}, suggesting that the results of the Zaklan model are robust.

An interesting extension of such models is to consider that the transition from honest to evader is not abrupt. In this case, it can be considered a third state that can be called susceptible of undecided \cite{davis,meu_econo}. The presence of such class was analyzed taking into account the dynamics of kinetic exchange opinion models \cite{meu_econo,biswas,meu,allan_pla}, and considering the same punishment rules of the Zaklan model. In this case, it was discussed \cite{meu_econo} that the presence of such third class affects substantially the dynamics, and that the compliance is high below the critical point (of the order-disorder transition) of the opinion dynamics governed by the kinetic exchanges. On the other hand, above the critical point the tax evasion can be considerably reduced by the enforcement mechanism.

In this work we propose another three-state agent-based model to analyze tax evasion dynamics. The transitions among the classes are ruled by probabilities, similarly to what happens in models of epidemic spreading \cite{bailey,anderson,satorras2015}. The enforcement mechanism is considering in the mentioned probabilities, as well as the social pressure of the contacts of a given individual. We will also see that the emergence of tax evaders in the population can be associated with a nonequilibrium phase transition, that was not observed in previous physics models of tax evasion, to the best of our knowledge \cite{zaklan,lima1,lima2,lima3,llacer,seibold,bertotti,meu_econo}.

This work is organized as follows. In section 2 we define the model's rules and the individuals presented in the population. After, we discuss results in three distinct topologies. Finally, in section 3 we present our conclusions and final remarks.


\section{Model and Results}

\qquad We considered a population of $N$ agents defined in a given network of contacts, that will be defined specifically in the following. Each individual $i$ ($i=1,2,...,N$) can be in one of three possible states or attitudes at a given time step $t$, represented by $X_{1}(t)$, $X_{2}(t)$ and $X_{3}(t)$. In other words, $X_{j}$ represents the number of individuals in a given state, with $j=1,2,3$. The state $X_{1}$ represents a \textit{honest tax payer}, i.e., an individual 100$\%$ convinced of his/her honesty, who does not consider evasion. He/she is either habitually compliant or he/she is a recent evader who has become honest as a result of enforcement efforts or social norms. On the other hand, the state $X_{3}$ represents a cheater, i.e, an individual who is an \textit{evading tax payer}. Whether a tax payer continues to evade depends on both enforcement and the effect of social interactions. Finally, the third state $X_{2}$ consists of taxpayers who are dissatisfied with the tax system (perhaps as a result of seeing others evade without being punished). These taxpayers are not actively evading, but they might if the perceived benefits of doing so exceed the perceived costs. For this group, evasion is an option, and so we classify them as \textit{susceptibles} \cite{davis}, i.e., they are susceptible to become evaders.

We consider here two distinct mechanisms to govern the transitions among the above-mentioned classes $X_{1}, X_{2}$ and $X_{3}$: social interactions and enforcement regime. The possible transitions are as follows:
\begin{eqnarray} \label{eq1}
X_{1} + X_{3} & \stackrel{\lambda}{\rightarrow} & X_{2} + X_{3} ~,  \\ \label{eq2}
X_{2} & \stackrel{\alpha}{\rightarrow} & X_{3} ~,  \\  \label{eq3}
X_{3} + X_{1} & \stackrel{\delta}{\rightarrow} & X_{1} + X_{1}  ~, \\ \label{eq4}
X_{3} & \stackrel{\beta}{\rightarrow} & X_{1} ~.
\end{eqnarray}

The interpretation of these transitions is as follows. Eq. (\ref{eq1}) represents an encounter of a honest agent $X_{1}$ with an evader $X_{3}$. In this case, with probability $\lambda$ the honest individual becomes susceptible $X_{2}$. The parameter $\lambda$ can be viewed as the persuasion power of the evaders. The following transition, Eq. (\ref{eq2}) represents a spontaneous transition from the susceptible state $X_{2}$ to the evader state $X_{3}$. The enforcement affects the behavior of a susceptible individual through its effect on the perceived costs of evasion (cost-benefit analysis). Thus, we assume that some susceptible tax payers will perceive that the benefits of evasion exceed the costs of evasion in each period, leading these individuals to evade. This is represented by the probability $\alpha$. In this case, we consider that the transition from honest to evader is not abrupt: the individual first becomes susceptible and after he might become evader.

Eq. (\ref{eq3}) represents the opposite transition in comparison with Eq. (\ref{eq1}). In this case, it represents an encounter of an evader agent $X_{3}$ with a honest tax payer $X_{1}$. In this case, with probability $\delta$ the evader individual becomes honest. The parameter $\delta$ can be viewed as the persuasion power of the honests. One can also consider that this last transition occurs to the state $X_{2}$, but for simplicity we consider that the evaders go directly to the honest compartment. 

Finnaly, Eq. (\ref{eq4}) represents another enforcement effect. We consider that
evaders become compliant after they are audited or when their perceptions regarding the costs and benefits of evasion change, either through experience or changing economic conditions \cite{davis}. This transition occurs with probability $\beta$, that can be viewed as a measure of the efficiency of the government's fiscalization. As in the previous case, one can also consider that some evaders might not be rehabilitated when they are audited, remaining susceptible rather than becoming honest, but for simplicity we will not consider those additional transitions.

In the following subsections we consider the model defined in Eqs. (\ref{eq1}) - (\ref{eq4}) on the top of three distinct topologies: fully-connected network, Erd\"{o}s-R\'enyi random graph and scale-free Barab\'asi-Albert network.


\subsection{Fully-connected network}

\qquad In this section we consider the model on a fully-connected graph. Considering the densities of each state, namely $x_{j}=X_{j}/N$ ($j=1,2,3$), the above Eqs. (\ref{eq1}) - (\ref{eq4}) can be translated on the mean-field equations
\begin{eqnarray} \label{eq5}
\frac{d}{dt}\,\,x_{1} & = & \beta\,x_{3} - \lambda\,x_{1}\,x_{3} + \delta\,x_{1}\,x_{3} ~, \\ \label{eq6}
\frac{d}{dt}\,x_{2} & = & -\alpha\,x_{2} + \lambda\,x_{1}\,x_{3} ~, \\ \label{eq7}
\frac{d}{dt}\,x_{3} & = & \alpha\,x_{2} - \beta\,x_{3} - \delta\,x_{1}\,x_{3} ~.
\end{eqnarray}
\noindent
where now $x_{1}$, $x_{2}$ and $x_{3}$ denote the fractions of honest, susceptible and tax evader individuals, respectively.

\begin{figure}[t]
\begin{center}
\vspace{6mm}
\includegraphics[width=0.33\textwidth,angle=270]{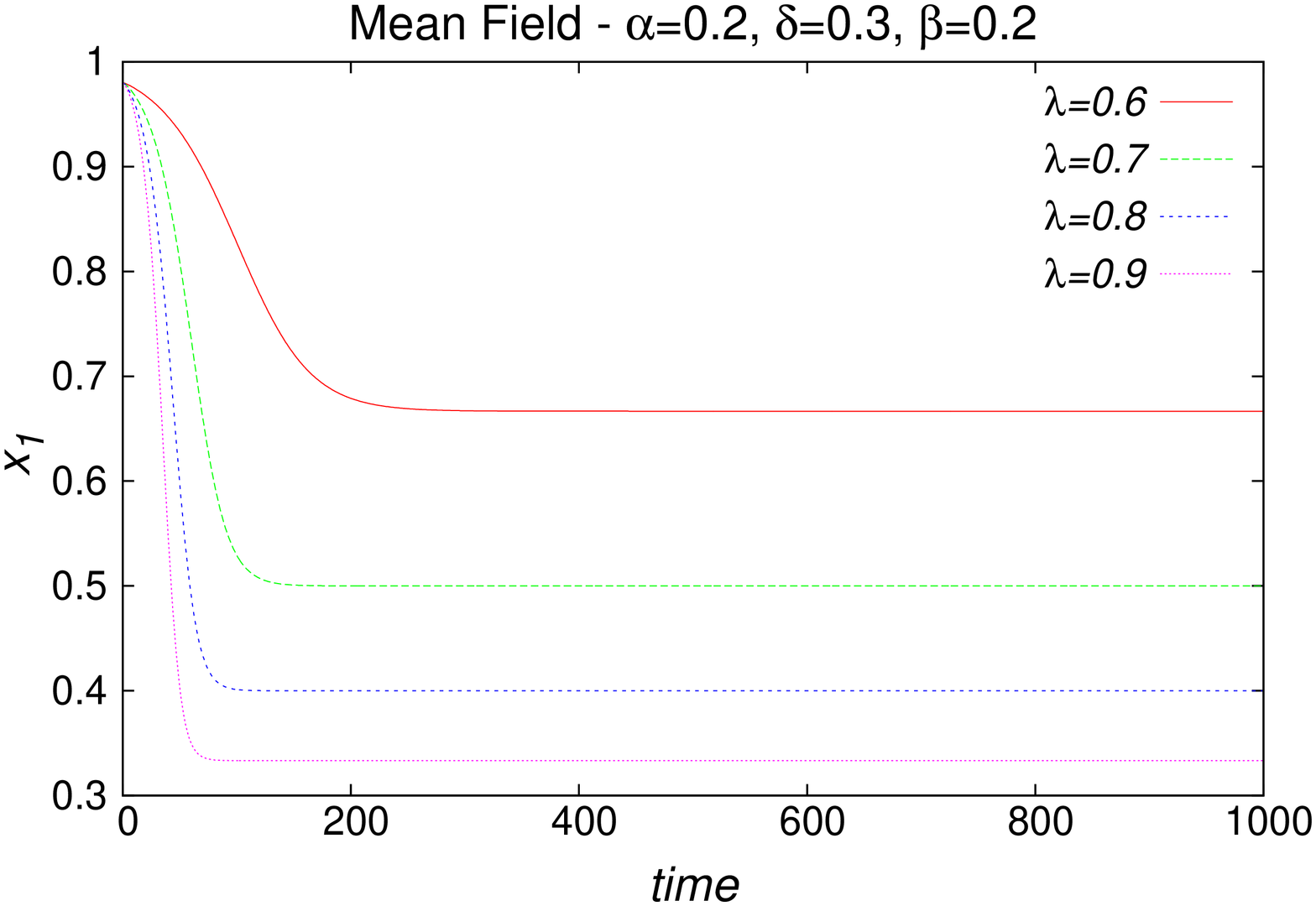}
\hspace{0.3cm}
\includegraphics[width=0.33\textwidth,angle=270]{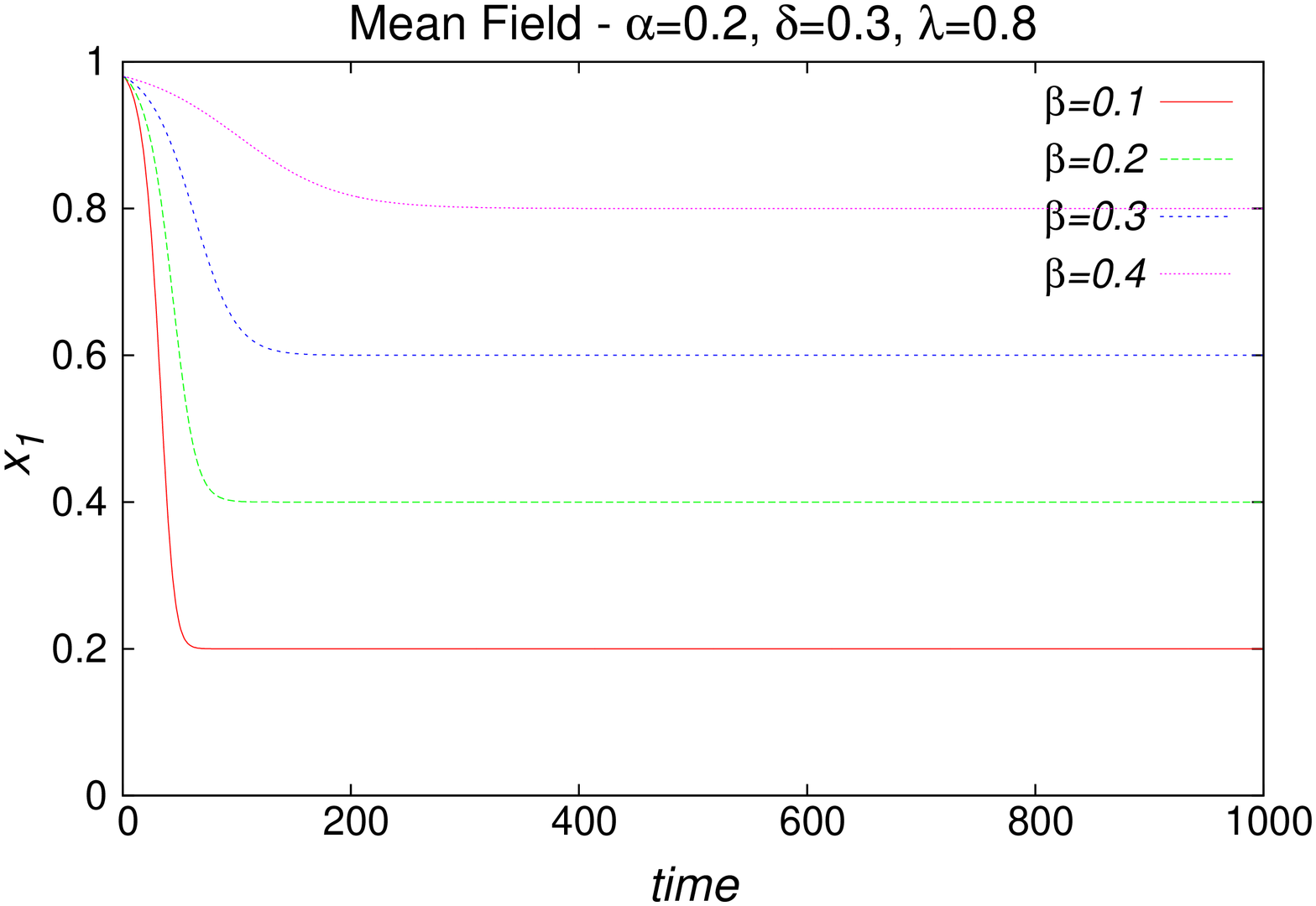}
\\
\vspace{0.5cm}
\includegraphics[width=0.33\textwidth,angle=270]{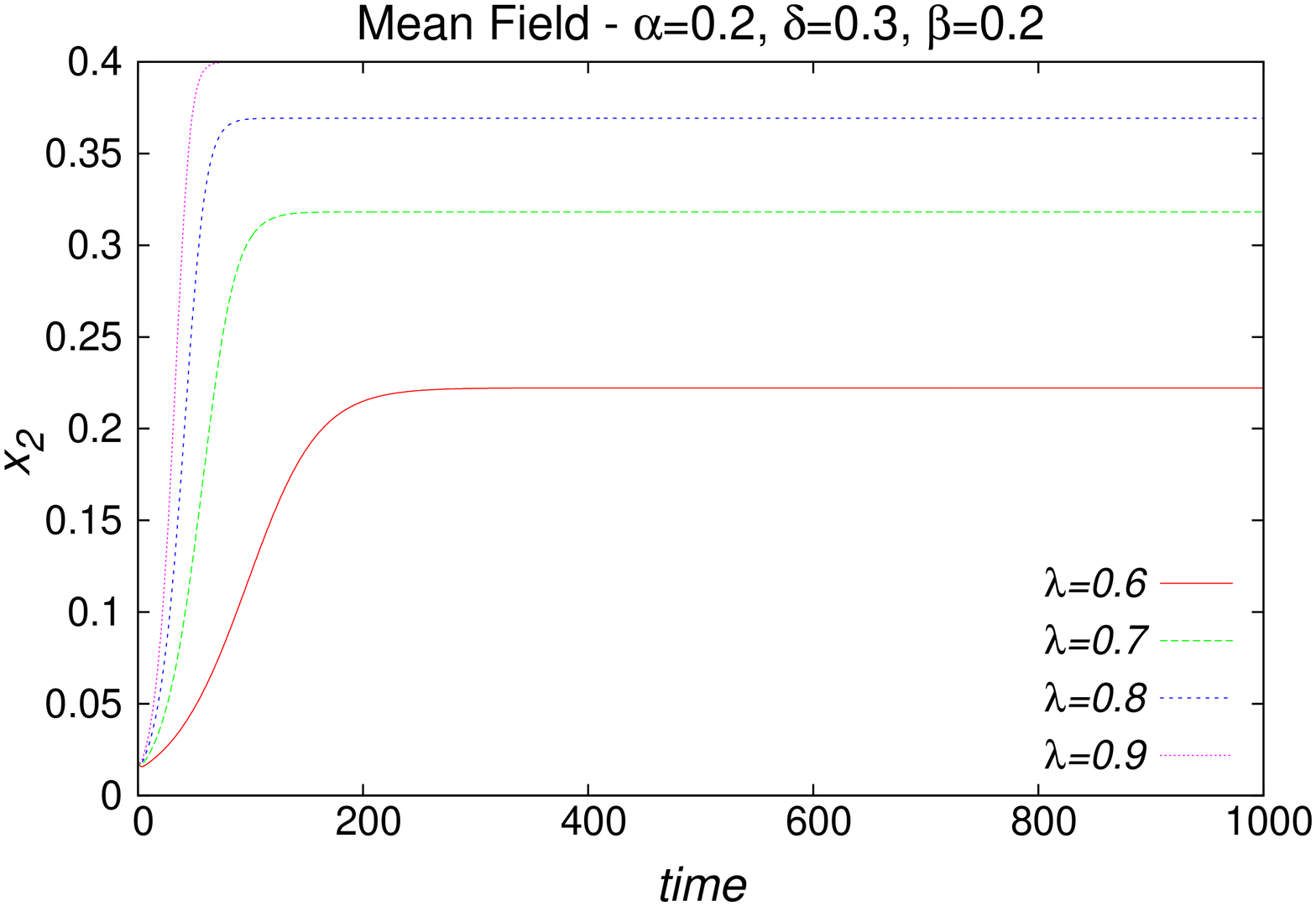}
\hspace{0.3cm}
\includegraphics[width=0.33\textwidth,angle=270]{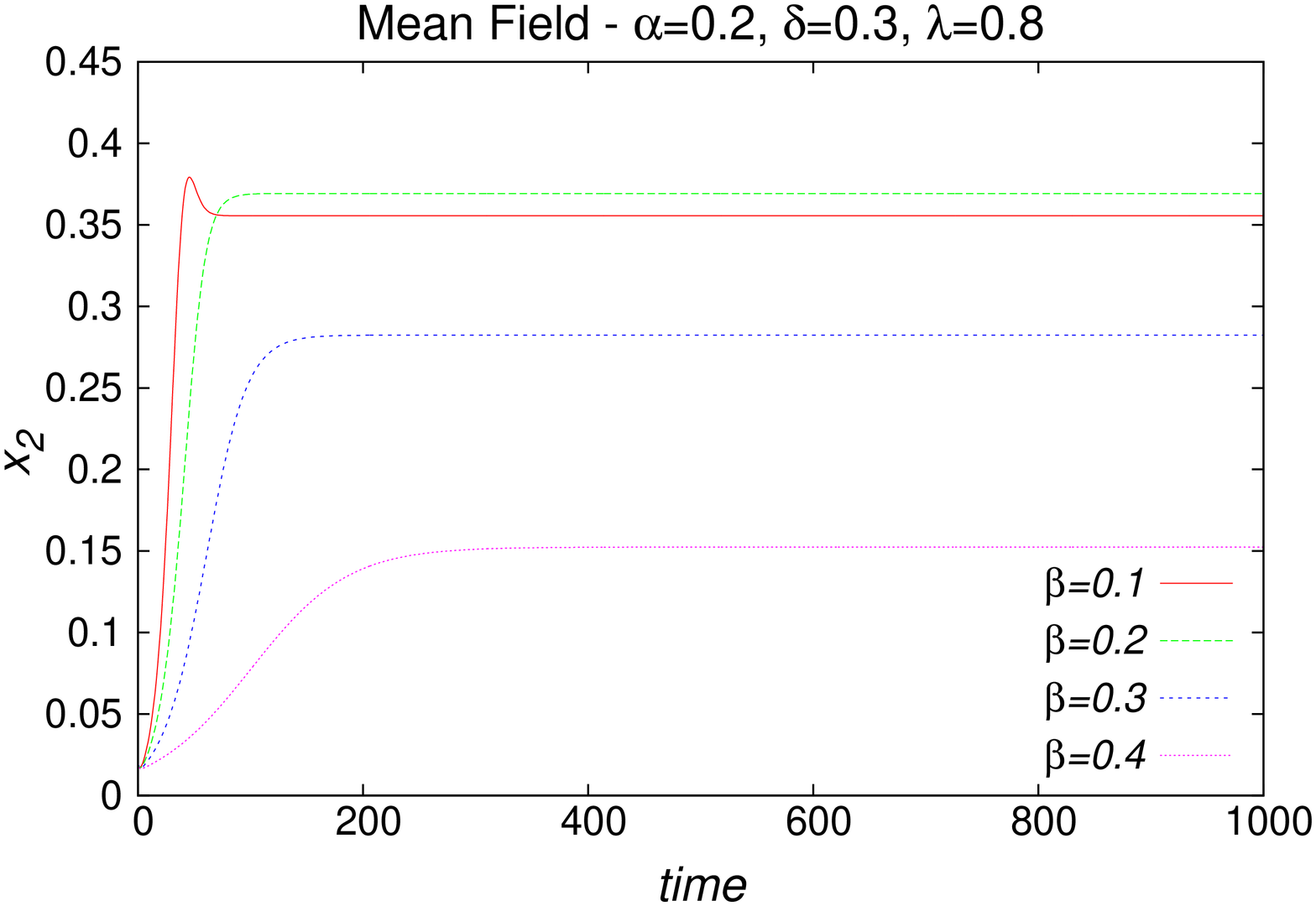}
\\
\vspace{0.5cm}
\includegraphics[width=0.33\textwidth,angle=270]{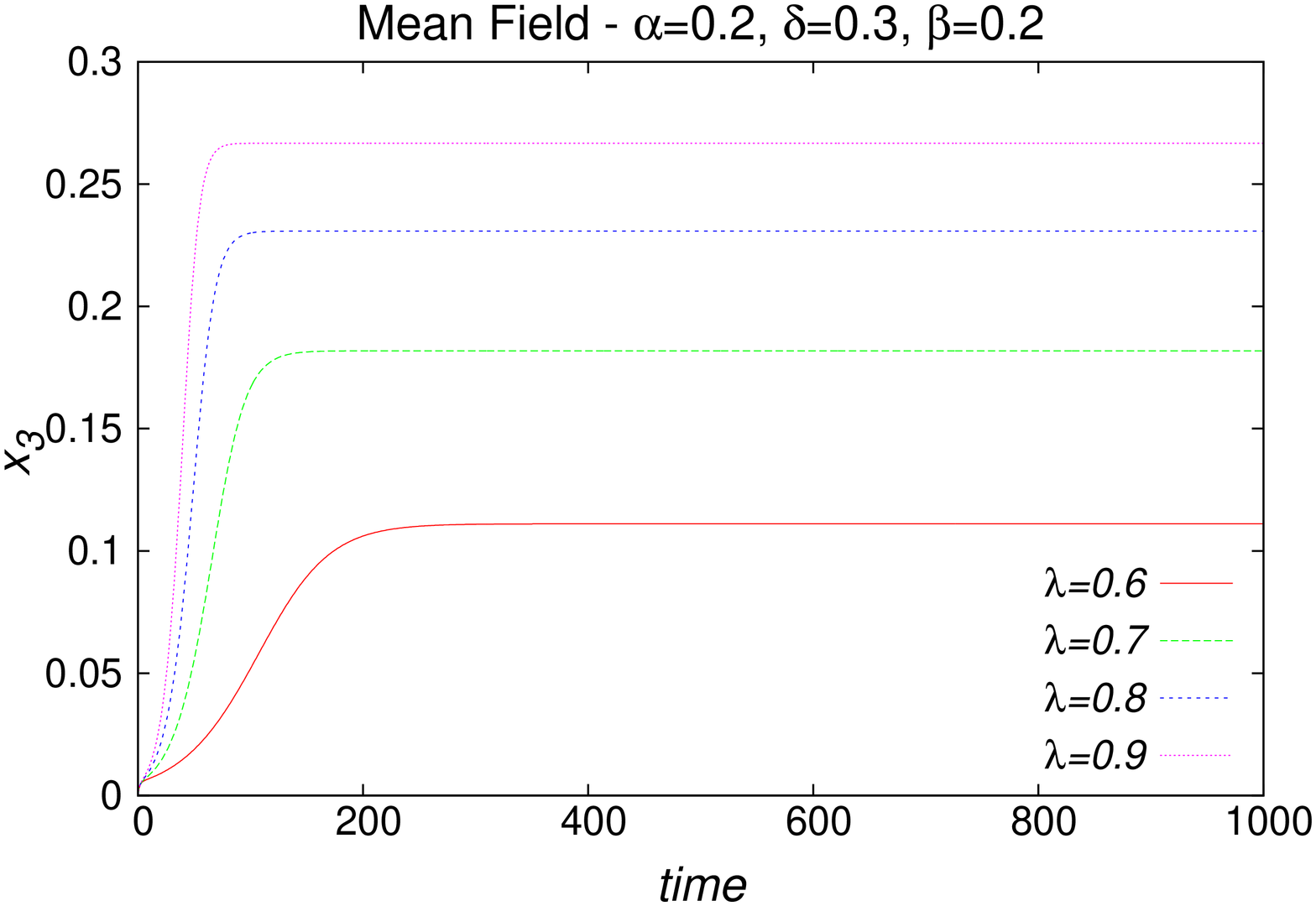}
\hspace{0.3cm}
\includegraphics[width=0.33\textwidth,angle=270]{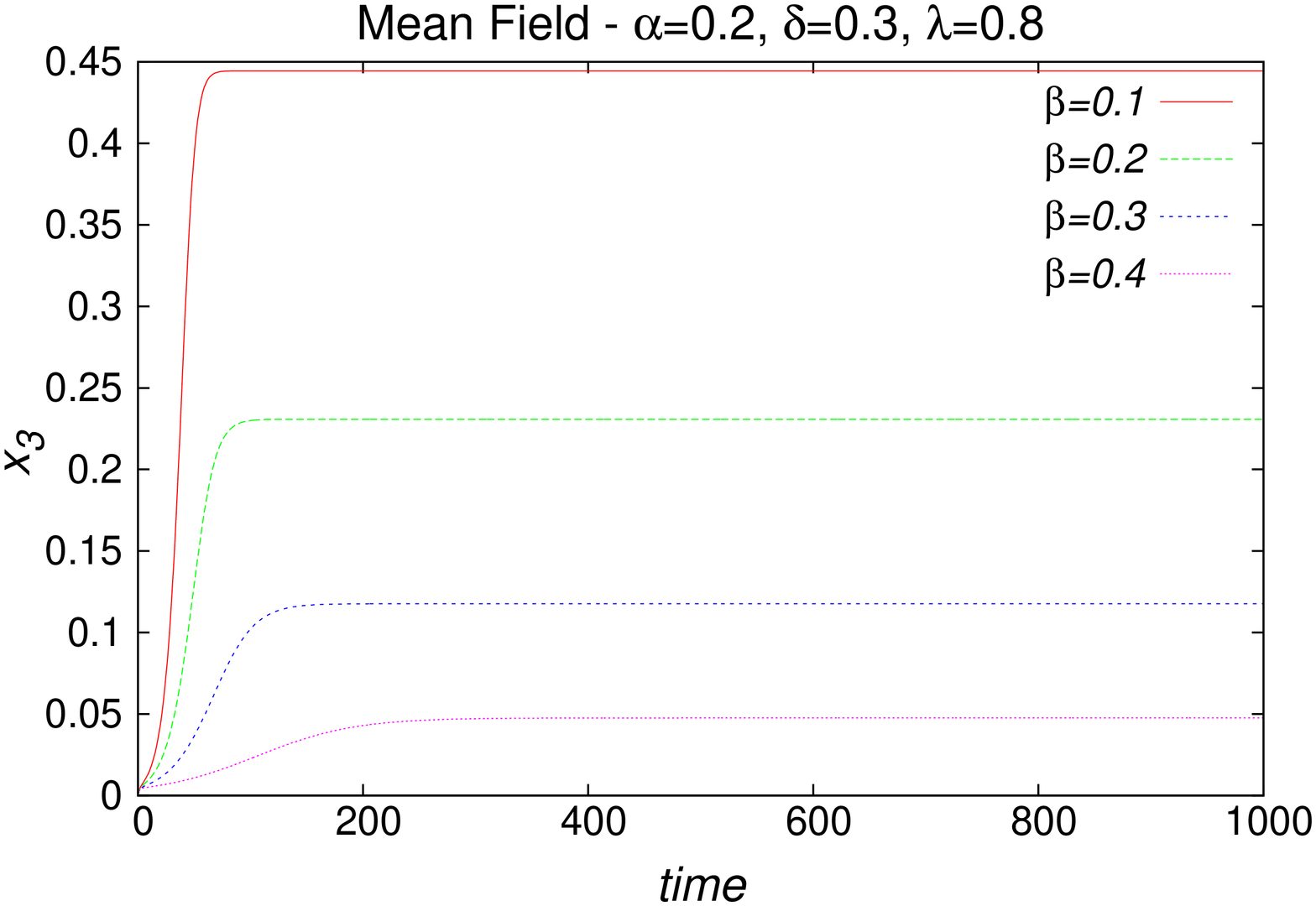}
\end{center}
\caption{(Color online) Time evolution of the three densities of agents $x_{1}$, $x_{2}$ and $x_{3}$ for the mean-field formulation of the model, based on Eqs. (\ref{eq5}) - (\ref{eq7}). The fixed parameters are $\alpha=0.2$ and $\delta=0.3$. In the left panels it is shown the evolution for $\beta=0.2$ and typical values of $\lambda$, whereas in the right panels we exhibit the evolution for $\lambda=0.8$ and typical values of $\beta$.}
\label{fig1}
\end{figure}

One can start analyzing the time evolution of the three classes of individuals. We numerically integrated Eqs. (\ref{eq5}), (\ref{eq6}) and (\ref{eq7}) in order to analyze the effects of the variation of the model's parameters. As initial conditions, we considered $x_{1}(0)=0.98$, $x_{2}(0)=0.02$ and $x_{3}(0)=0$, and for simplicity we fixed $\alpha=0.2$ and $\delta=0.3$, varying the parameters $\lambda$ and $\beta$. In Fig. \ref{fig1} we exhibit results for fixed $\beta=0.2$ and typical values of $\lambda$ (left panels) and for fixed $\lambda=0.8$ and typical values of $\beta$ (right panels). For the cases with fixed $\beta$, one can see that the increase of $\lambda$ causes the decrease of $x_{1}$ and the increase of $x_{2}$ and $x_{3}$. Remembering that $\lambda$ models the persuasion of evaders $x_{3}$ in the social interactions with honests $x_{1}$, i.e. the transition given by eq. (\ref{eq1}), it is easier to understand those results: for increasing values of $\lambda$ more agents goes from the class $x_{1}$ to the class $x_{2}$, and these susceptible individuals after can go to the evader class $x_{3}$, which cause an increse of the susceptible and evader classes, and the decrease of honests.

On the other hand, in the graphics with fixed $\lambda$, one can see that the increase of $\beta$ leads to an increase of honests and a decrease of susceptibles and evaders. This is compatible with the interpretation of $\beta$ as a government fiscalization: the increase of the efficiency of the enforcement regime leads to the rise of honesty in the population, as well as the decrease of tax evasion. Thus, the variation of those two parameters $\beta$ and $\lambda$ models the competition of pressure of the social contacts and the State's fiscalization.

One can observe in Fig. \ref{fig1} that the fractions $x_{1}$, $x_{2}$ and $x_{3}$ evolve with time and after some steps they stabilize. One can derive analytically the stationary fractions of the three classes by taking the time derivatives equal to zero in Eqs. (\ref{eq5}), (\ref{eq6}) and (\ref{eq7}). In this case, one can obtain the fixed points as functions of the models'parameters,
\begin{eqnarray} \label{eq8}
x_{1}^{*} & = & \frac{\beta}{\lambda-\delta}  ~, \\ \label{eq9}
x_{2}^{*} & = & \frac{\lambda\,\beta\,(\lambda-\delta-\beta)}{(\lambda-\delta)\,[\lambda\,\beta+\alpha\,(\lambda-\delta)]}  ~, \\ \label{eq10}
x_{3}^{*} & = & \frac{\lambda-\delta-\beta}{\lambda-\delta+(\lambda/\alpha)\,\beta}  ~.
\end{eqnarray}
\begin{figure}[t]
\begin{center}
\vspace{6mm}
\includegraphics[width=0.33\textwidth,angle=270]{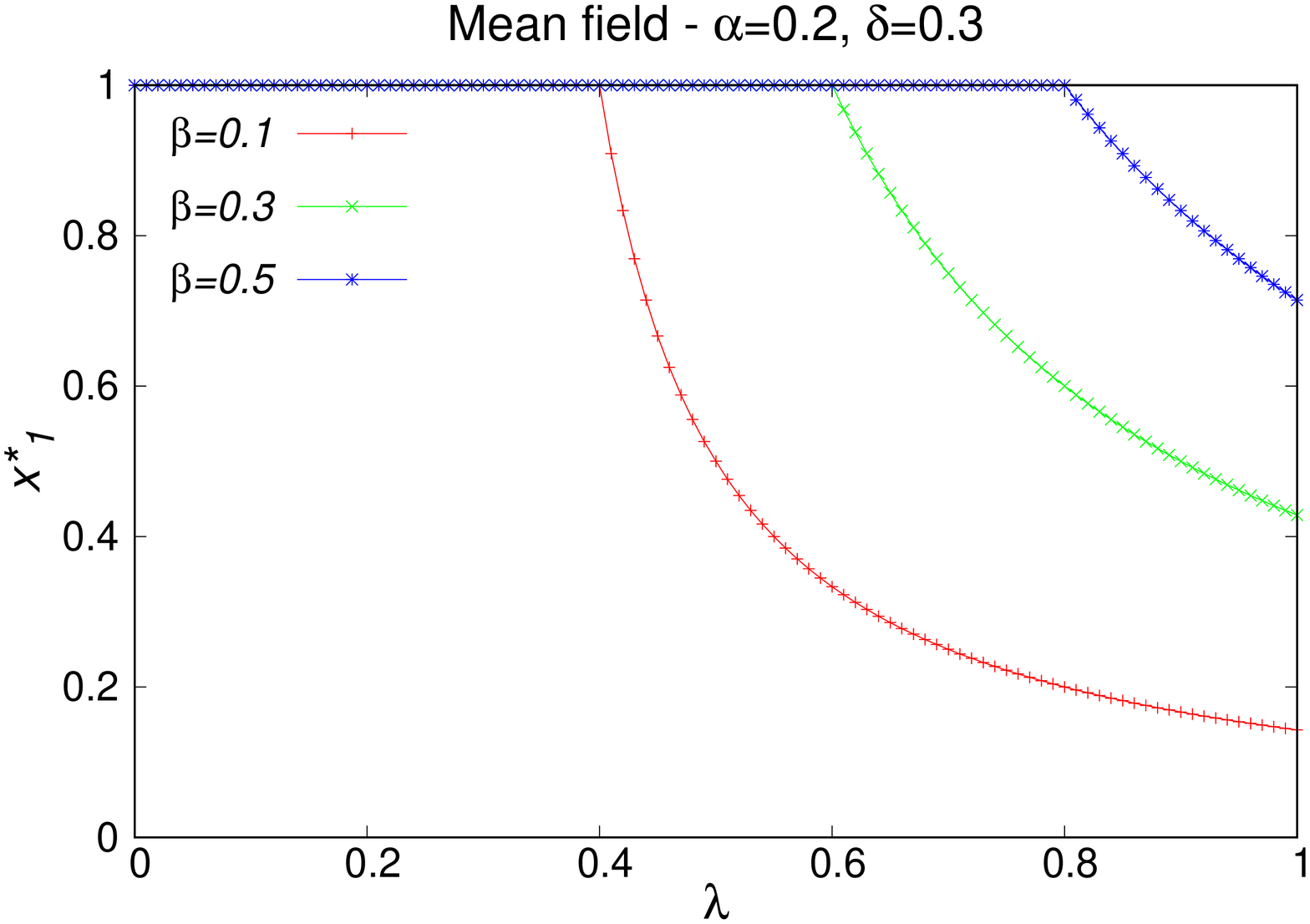}
\hspace{0.3cm}
\includegraphics[width=0.33\textwidth,angle=270]{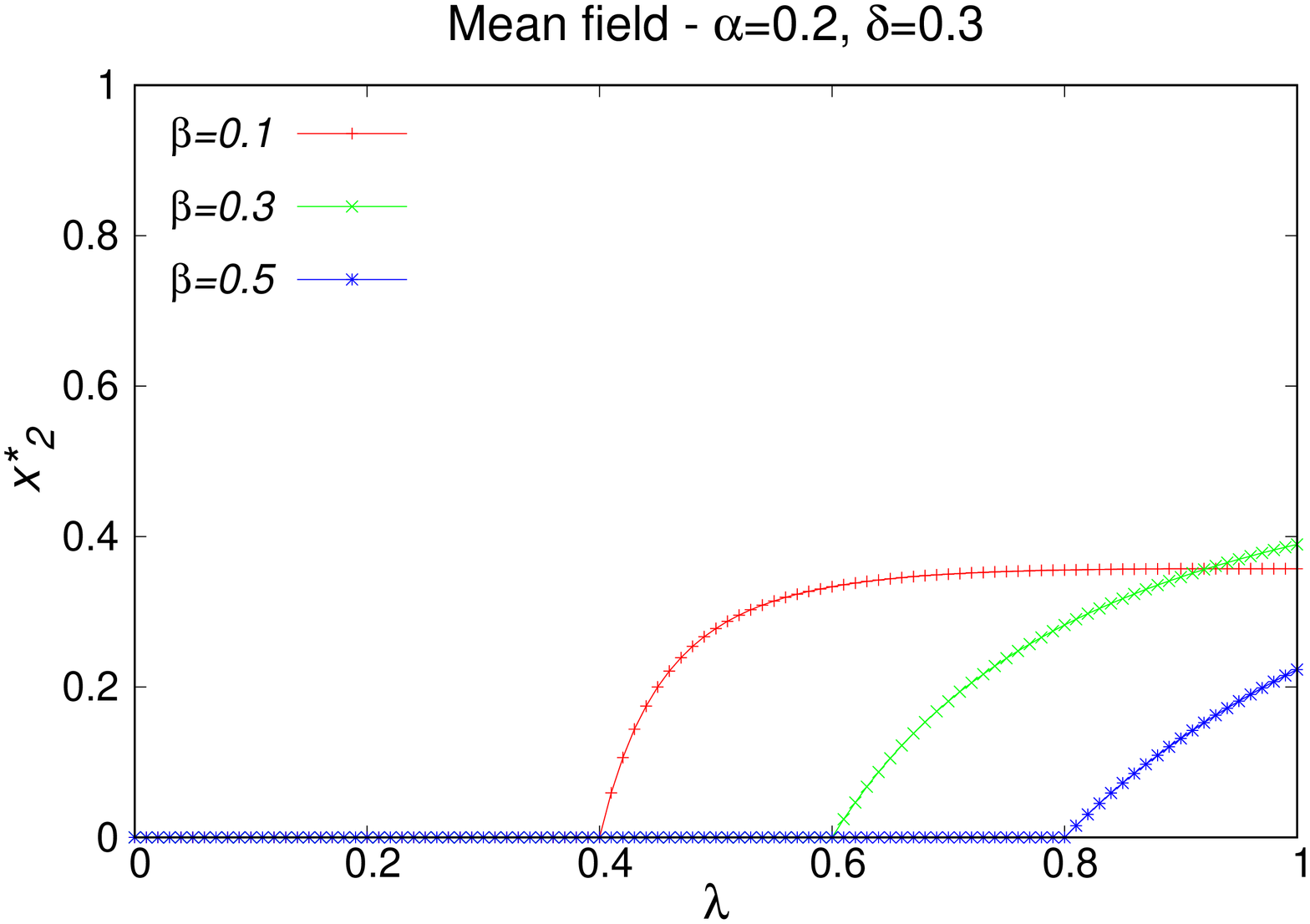}
\\
\vspace{0.5cm}
\includegraphics[width=0.33\textwidth,angle=270]{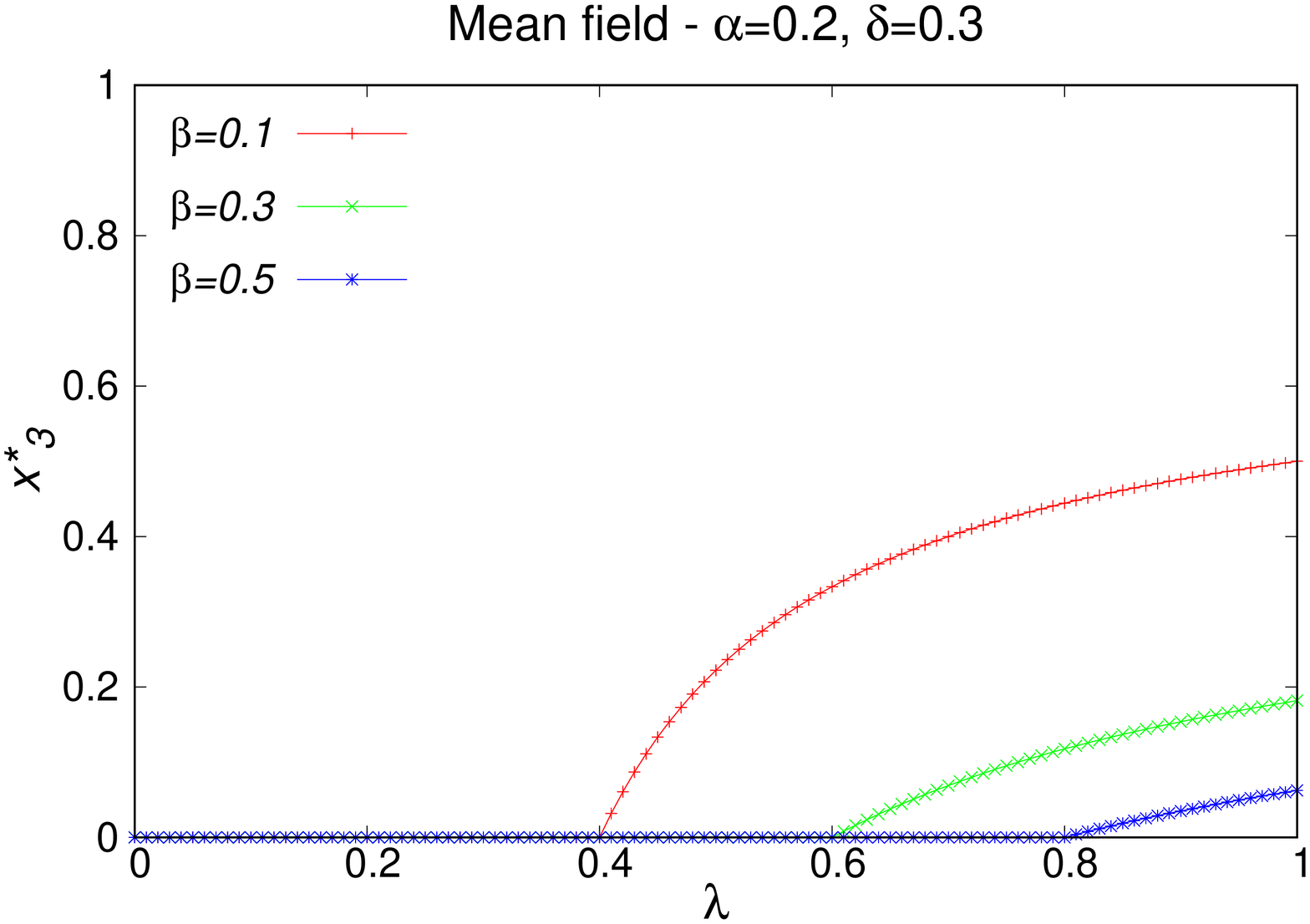}
\end{center}
\caption{(Color online) Stationary fractions $x_{1}^{*}$ (upper left), $x_{2}^{*}$ (upper right) and $x_{3}^{*}$ (lower) of the three classes of agents for the mean-field formulation of the model, given by Eqs. (\ref{eq8}) - (\ref{eq10}). The fractions are plotted as a function of $\lambda$ for typical values of $\beta$. The fixed parameters are $\alpha=0.2$ and $\delta=0.3$.}
\label{fig2}
\end{figure}

\noindent
One can see from the above equations that the model undergoes a nonequilibrium phase transition if we consider the stationary fraction of evaders $x_{3}^{*}$ as an order parameter: for $\lambda\leq\lambda_{c}$ the stationary solutions are given by $x_{1}^{*}=1$ and $x_{2}^{*}=x_{3}^{*}=0$, whereas for $\lambda>\lambda_{c}$ the solutions give us $x_{1}^{*}>0$, $x_{2}^{*}>0$ and $x_{3}^{*}>0$, where the threshold is given by $\lambda_{c}=\beta+\delta$. This is an active-absorbing transition \cite{dickman,hinrichsen}, and it separates a phase where the tax evaders disappear of the population in the long-time limit and the population is formed only by honests, from a phase where there is a finite fraction of evaders in the long time. The susceptible agents also survive in the active phase, and they disappear in the absorbing phase.

For a better analysis of the stationary behavior, we show in Fig. \ref{fig2} the stationary fractions $x_{1}^{*}$, $x_{2}^{*}$ and $x_{3}^{*}$ as functions of $\lambda$ for typical values of $\beta$. The results are based on Eqs. (\ref{eq8}) - (\ref{eq10}). One can see the mentioned phase transition in the lower panel: for values $\lambda\leq \lambda_{c}=\beta+\delta$ we have $x_{3}^{*}=0$, and for $\lambda>\lambda_{c}$ we have $x_{3}^{*}>0$. In addition, one can see again the behaviors discussed above, i.e., the decrease of the evasion for increasing fiscalization ($\beta$) and the decrease of honests due to social pressure of tax evaders ($\lambda$), that are realistic features of the model. Comparing the three values of $\beta$ in the lower panel of Fig. \ref{fig2}, we see that the enforcement regime can be extremely effective for control the evasion. Indeed, this effect can be seen in Eqs.  (\ref{eq8}) - (\ref{eq10}): for increasing $\beta$ we see that $x_{3}^{*}$ decreases, and $x_{1}^{*}$ increases.


\subsection{Erd\"{o}s-R\'enyi Random graph}

\qquad In this section we consider the model on Erd\"{o}s-R\'enyi (ER) random graphs. The network is formed by $N$ isolated nodes, and we connect each pair with probability $p$. In this case, we performed simulations considering the rules given by Eqs. (\ref{eq1})-(\ref{eq4}) for network size $N=10^{4}$ and the connection probability $p=5\times 10^{-4}$, which gives us an average connectivity $\langle k\rangle=5$. 

The numerical procedure is as follows. We visit every node in the ER graph and apply the rules (\ref{eq1})-(\ref{eq4}). In the case where the chosen node is in the $X_{1}$ state, for example, we apply the rule (\ref{eq1}) if he/she has at least one neighbor in the $X_{3}$ state. The same occurs for the social interaction given by Eq. (\ref{eq3}). The remaining rules (\ref{eq2}) and (\ref{eq4}) are spontaneous transitions.

\begin{figure}[t]
\begin{center}
\vspace{6mm}
\includegraphics[width=0.33\textwidth,angle=270]{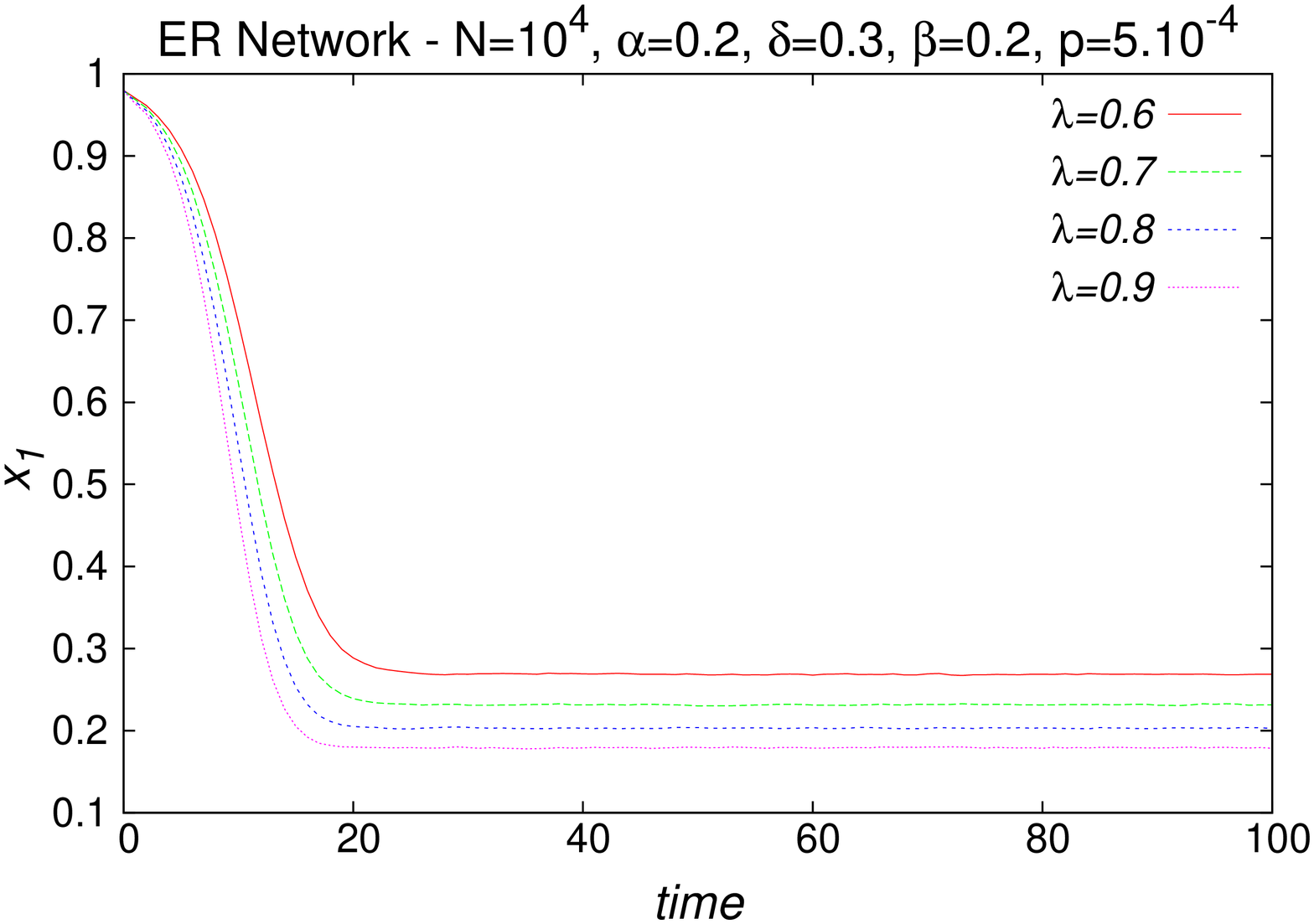}
\hspace{0.3cm}
\includegraphics[width=0.33\textwidth,angle=270]{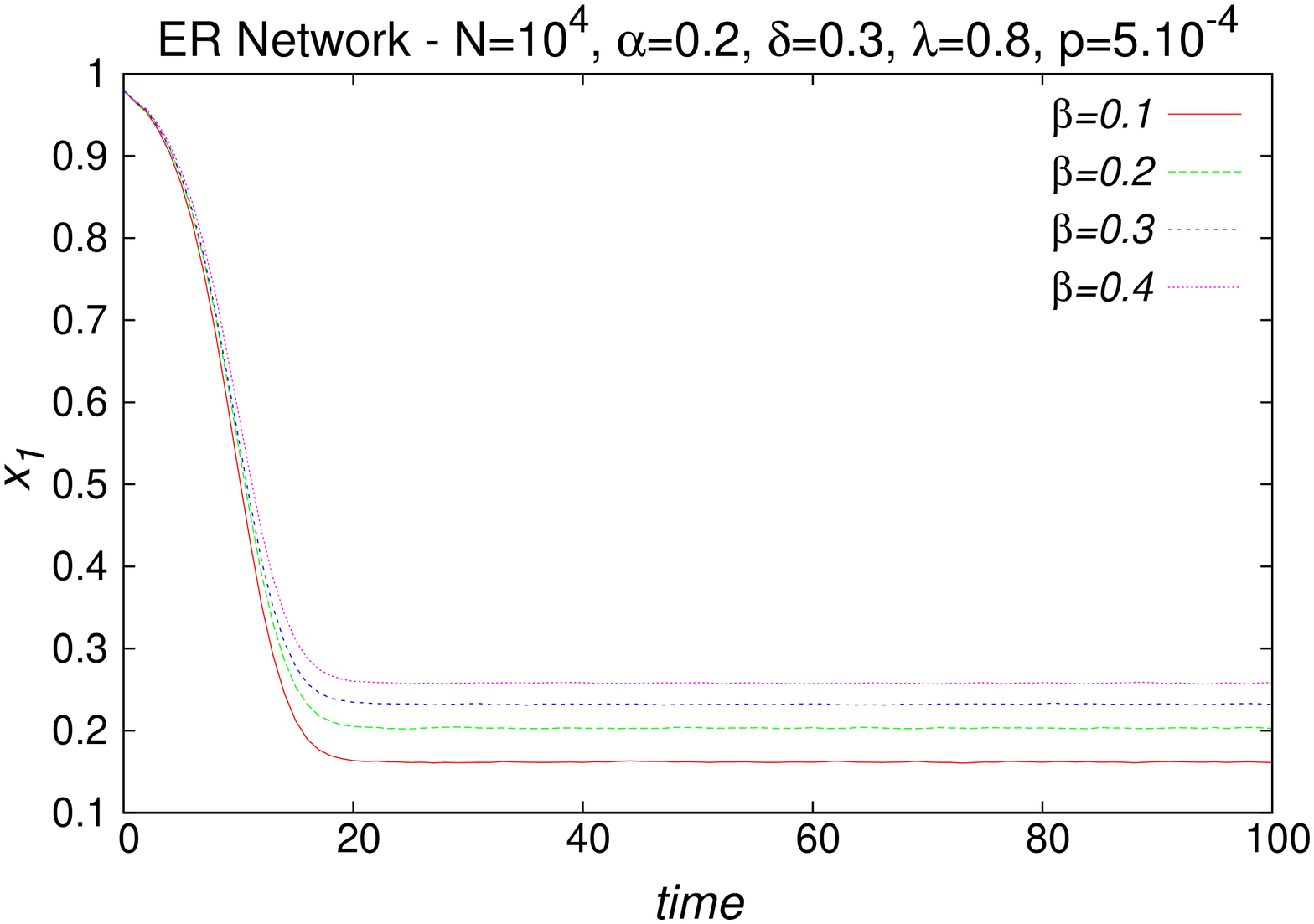}
\\
\vspace{0.5cm}
\includegraphics[width=0.33\textwidth,angle=270]{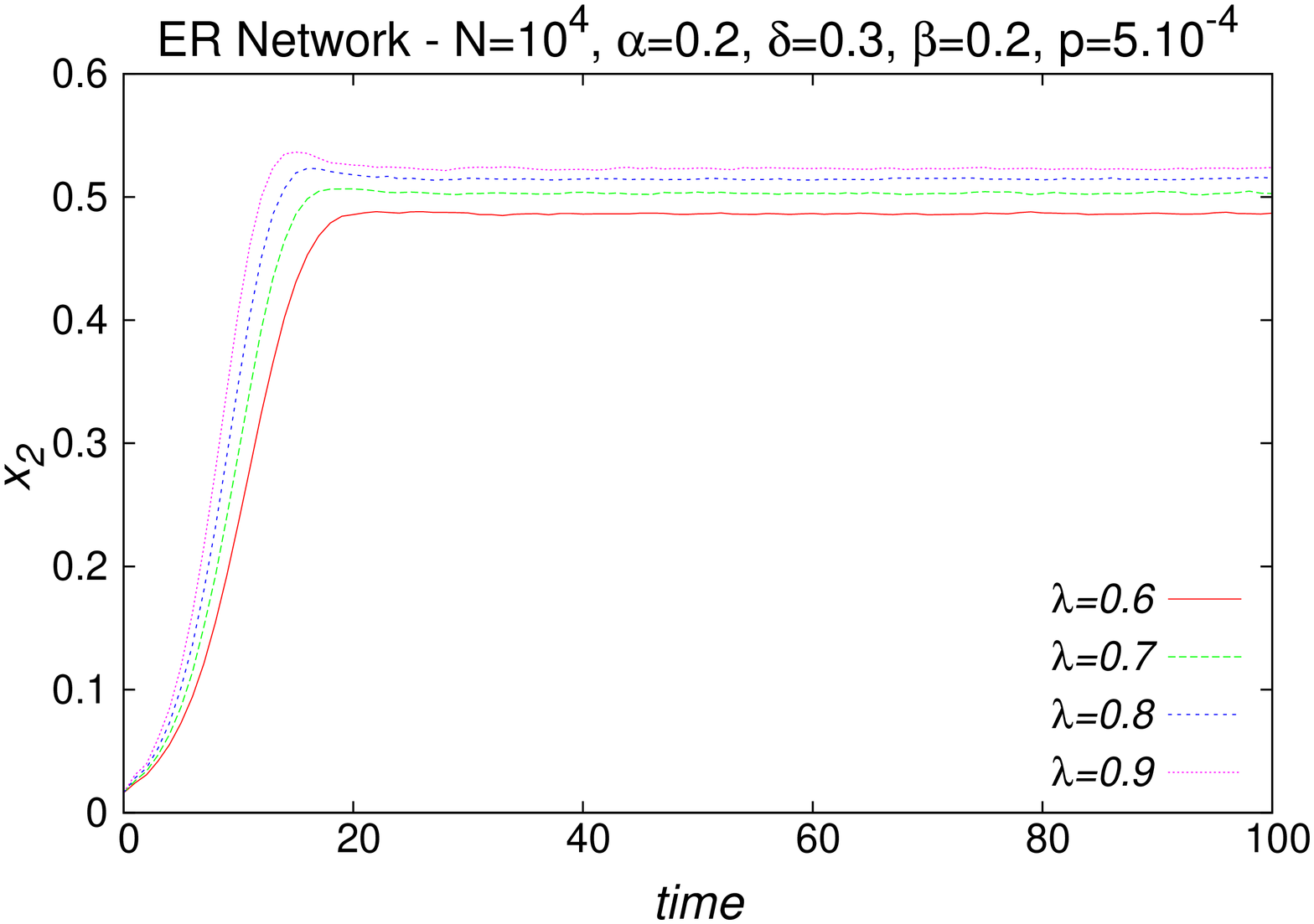}
\hspace{0.3cm}
\includegraphics[width=0.33\textwidth,angle=270]{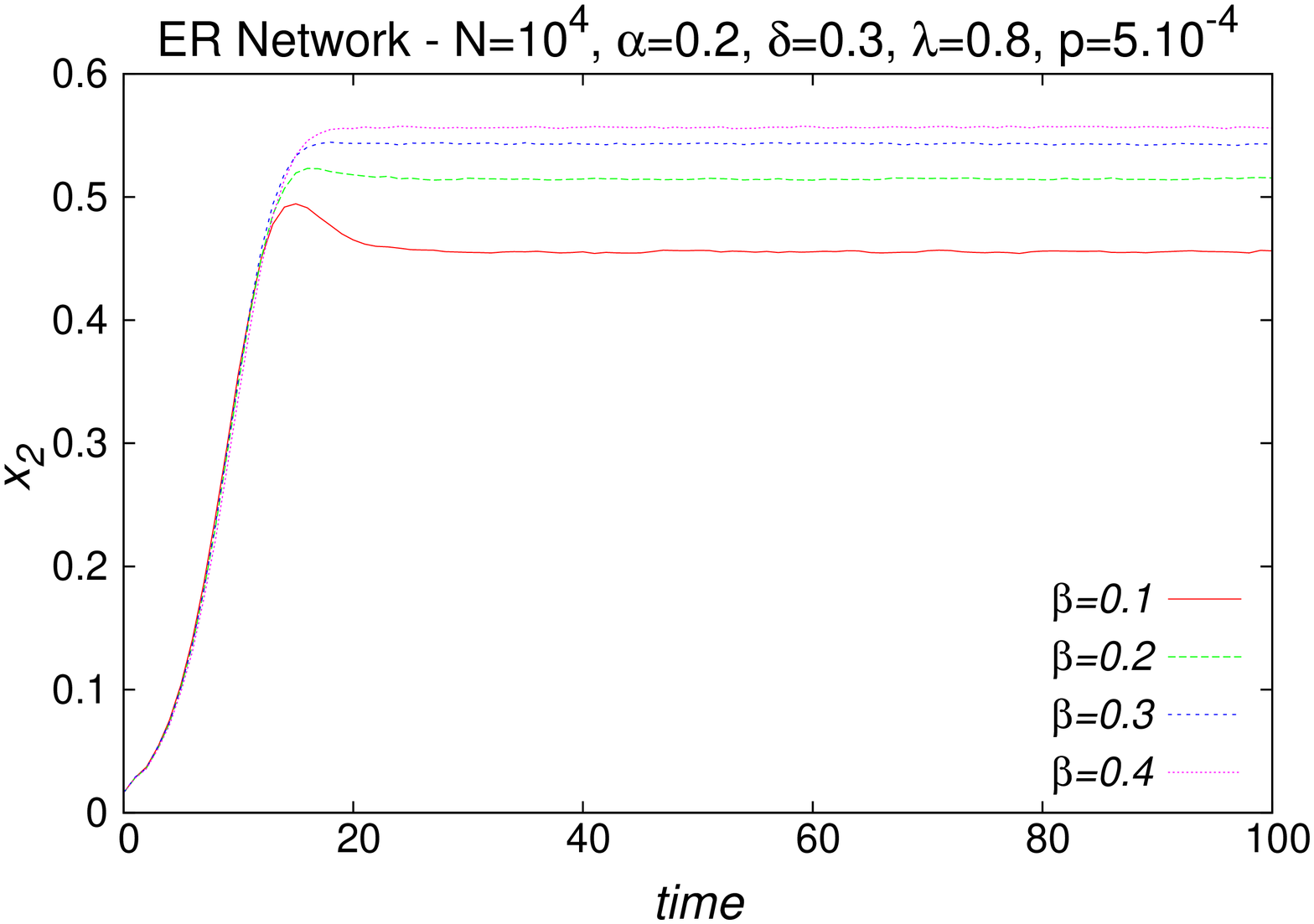}
\\
\vspace{0.5cm}
\includegraphics[width=0.33\textwidth,angle=270]{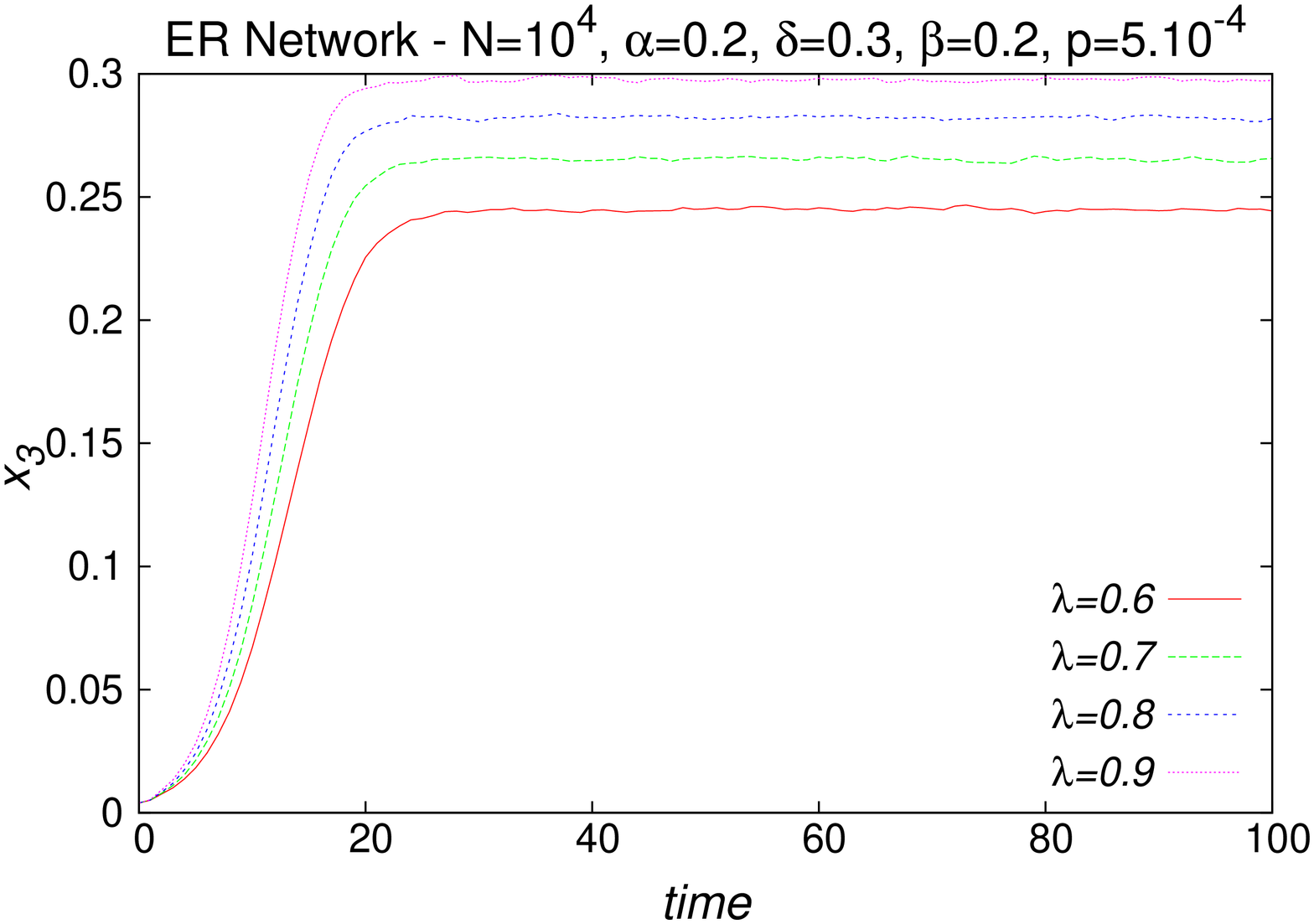}
\hspace{0.3cm}
\includegraphics[width=0.33\textwidth,angle=270]{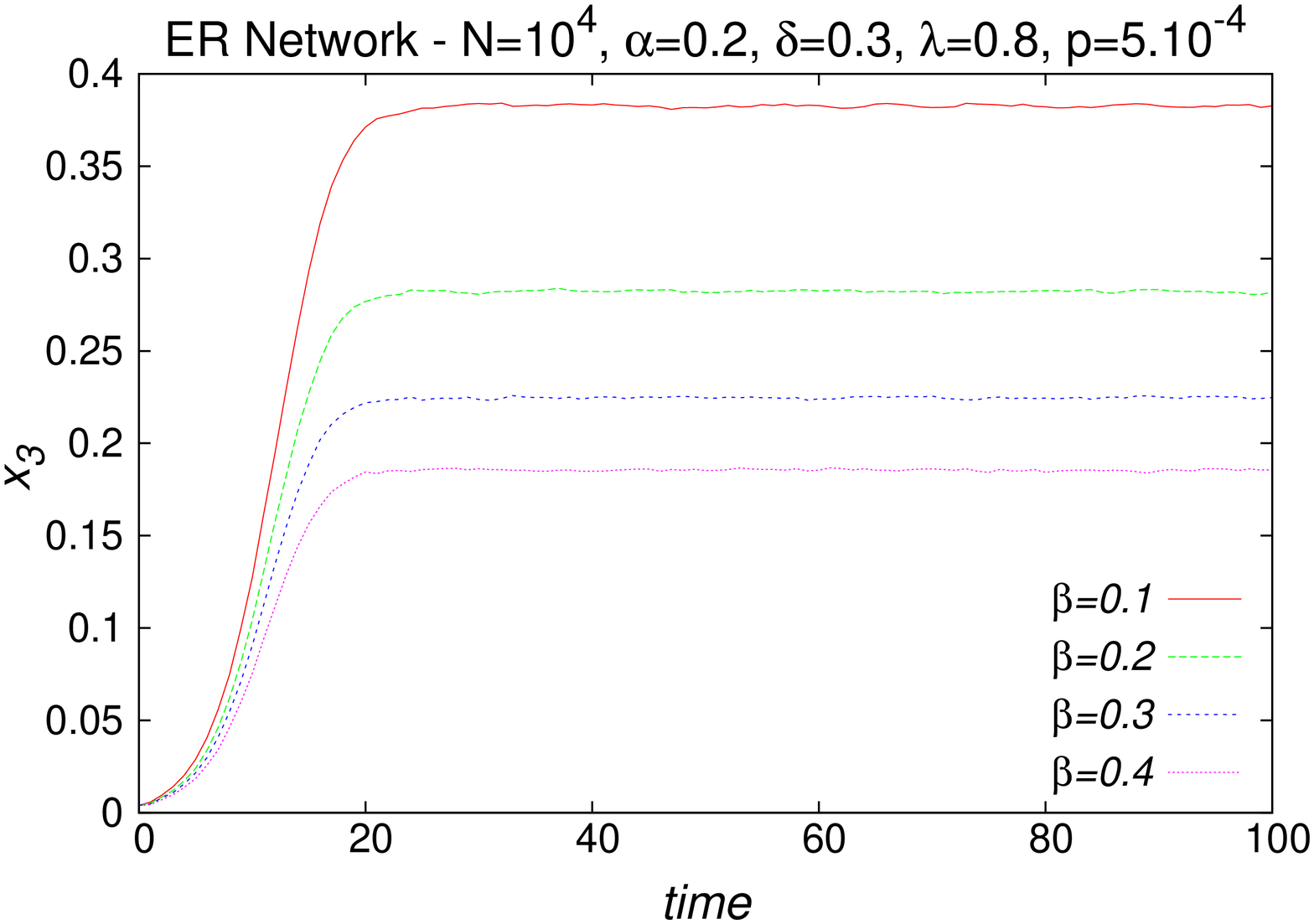}
\end{center}
\caption{(Color online) Time evolution of the three densities of agents $x_{1}$, $x_{2}$ and $x_{3}$ obtained from simulations of the model defined on the ER Random Graph. The fixed parameters are $\alpha=0.2$ and $\delta=0.3$. In the left panels it is shown the evolution for $\beta=0.2$ and typical values of $\lambda$, whereas in the right panels we exhibit the evolution for $\lambda=0.8$ and typical values of $\beta$.}
\label{fig3}
\end{figure}

As in the previous subsection, one can start analyzing the time evolution of the three classes of individuals. We considered the same initial conditions as before, namely $x_{1}(0)=0.98$, $x_{2}(0)=0.02$ and $x_{3}(0)=0$, and for simplicity we fixed $\alpha=0.2$ and $\delta=0.3$, varying the parameters $\lambda$ and $\beta$. In Fig. \ref{fig3} we exhibit results for fixed $\beta=0.2$ and typical values of $\lambda$ (left panels) and for fixed $\lambda=0.8$ and typical values of $\beta$ (right panels). One can see a qualitative similar behavior observed in the fully-connected graph, i.e., for the cases with fixed $\beta$, one can see that the increase of $\lambda$ leads to the decrease of $x_{1}$ and the increase of $x_{2}$ and $x_{3}$, since $\lambda$ is related to the social pressure of tax evaders over honest individuals. In addition, for the graphics with fixed $\lambda$, the increase of the fiscalization $\beta$ leads to an increase of honests and a decrease of susceptibles and evaders. 

\begin{figure}[t]
\begin{center}
\vspace{6mm}
\includegraphics[width=0.33\textwidth,angle=270]{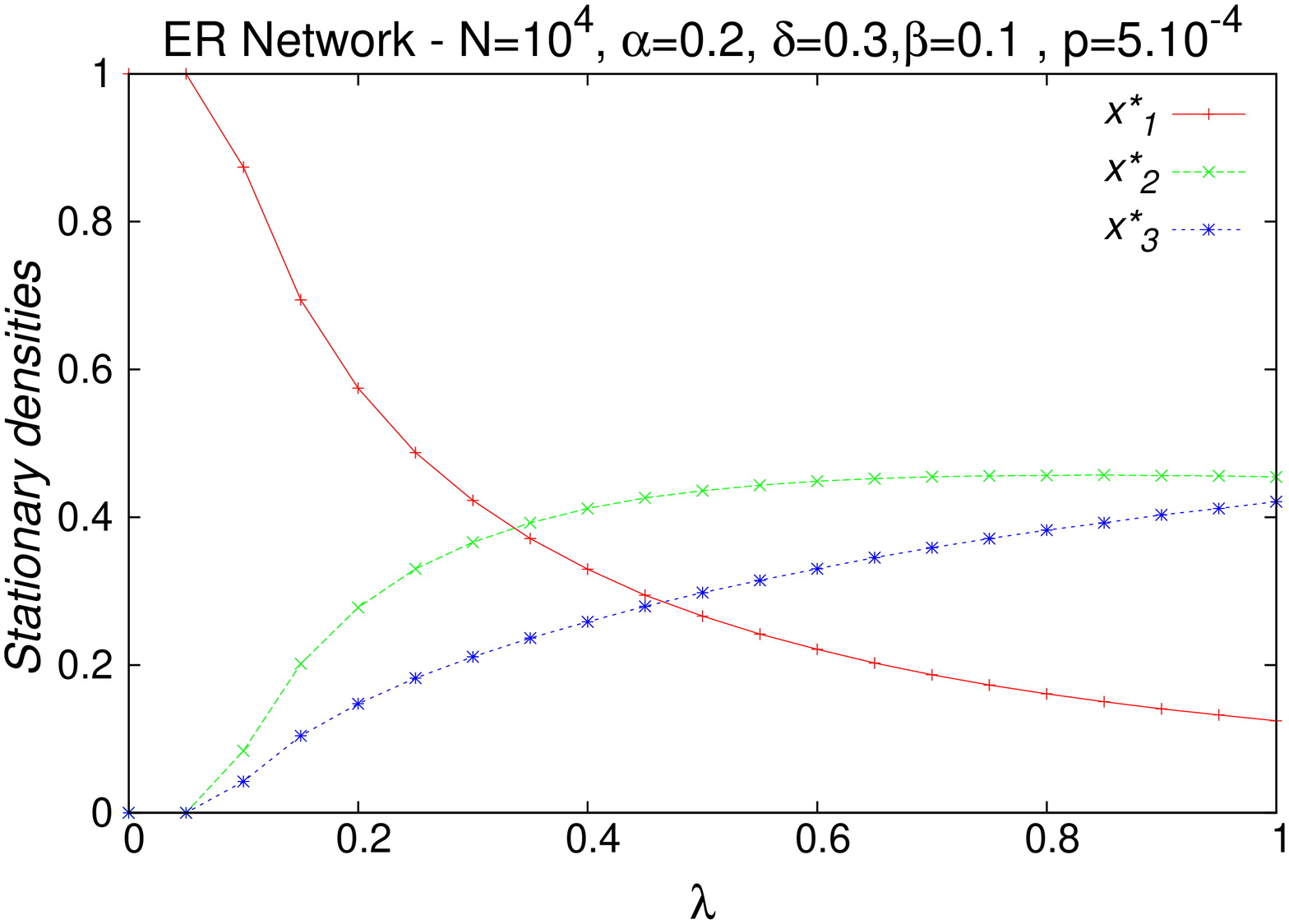}
\hspace{0.3cm}
\includegraphics[width=0.33\textwidth,angle=270]{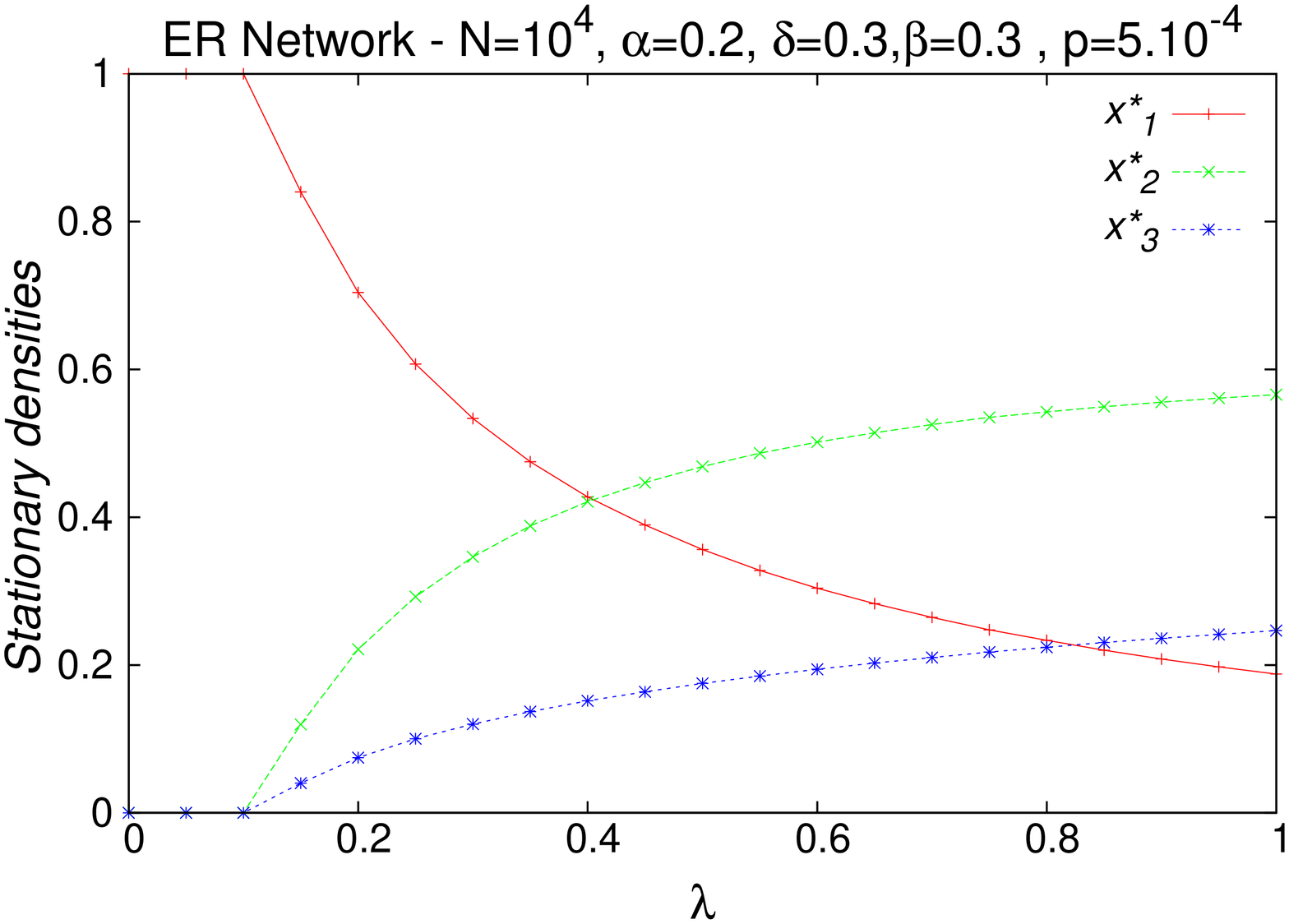}
\\
\vspace{0.5cm}
\includegraphics[width=0.33\textwidth,angle=270]{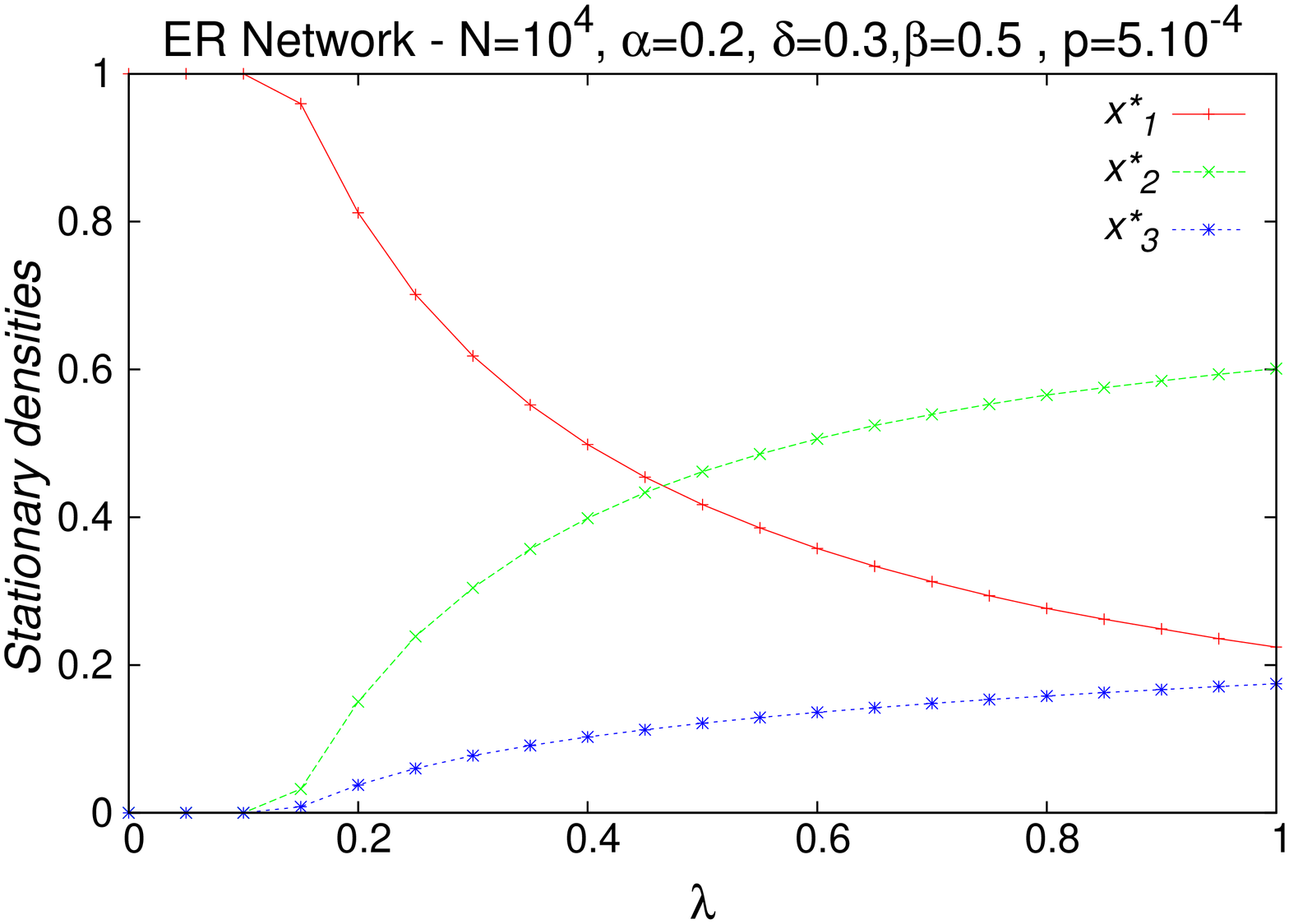}
\end{center}
\caption{(Color online) Stationary fractions $x_{1}^{*}$, $x_{2}^{*}$ and $x_{3}^{*}$ as functions of $\lambda$ for $\beta=0.1$ (upper left), $\beta=0.3$ (upper right) and $\beta=0.5$ (lower) for the model simulated on the ER graph. The fixed parameters are $\alpha=0.2$ and $\delta=0.3$.}
\label{fig4}
\end{figure}

However, the stationary values are different from the previous cases, as well as the impact of fiscalization and social pressure. In order to better see these differences, we exhibit in Fig. \ref{fig4} the stationary values $x_{1}^{*}$, $x_{2}^{*}$ and $x_{3}^{*}$ as functions of $\lambda$ for typical values of the fiscalization $\beta$. Comparing the three graphics, one can see that the fiscalization can reduce the fraction of tax evaders in the population, even if the social pressure $\lambda$ of dishonest individuals over honest ones is high: if $\beta$ is increased from $0.1$ to $0.5$ the stationary fraction of evaders in the population reduces from $\approx 0.4$ to $\approx 0.2$ for $\lambda=1.0$. However, in comparison with the mean-field case, the reduction of evasion is smaller. Thus, considering a more realistic topology, the pressure of the social contacts in the network leads to a slow decrease of the honest tax payers in comparison with the fully-connected graph. This occurs since a given agent in the ER random graph is always connected with the same neighbors (average value $\langle k\rangle$), and in the fully-connected case each agent can interact with all others.

Furthermore, one can see a large density of susceptibles in comparison with the fully-connected graph. One can also see the above-mentioned active-absorbing phase transition, but the threshold values are very small in comparison with the mean-field case. All these differences appear as consequences of the presence of a more realistic topology for modelling the society.

\subsection{Barab\'asi-Albert network}

\qquad Finally, in this section we consider the model on Barab\'asi-Albert (BA) scale-free networks. In this case, we performed simulations considering the rules given by Eqs. (\ref{eq1})-(\ref{eq4}) for network size $N=10^{4}$. Each generated network starts with $2$ nodes connected between themselves, and at each time step we add 1 node and 1 link to a pre-existing node, considering the usual preferential attachment procedure (probability proportional to the connectivity). The numerical procedure is the same described for the ER graph: we visit every node in the BA network and apply the rules (\ref{eq1})-(\ref{eq4}). In the case where the chosen node is in the $X_{1}$ state, for example, we apply the rule (\ref{eq1}) if he/she has at least one neighbor in the $X_{3}$ state. The same occurs for the social interaction given by Eq. (\ref{eq3}). The remaining rules (\ref{eq2}) and (\ref{eq4}) are spontaneous transitions.

\begin{figure}[t]
\begin{center}
\vspace{6mm}
\includegraphics[width=0.33\textwidth,angle=270]{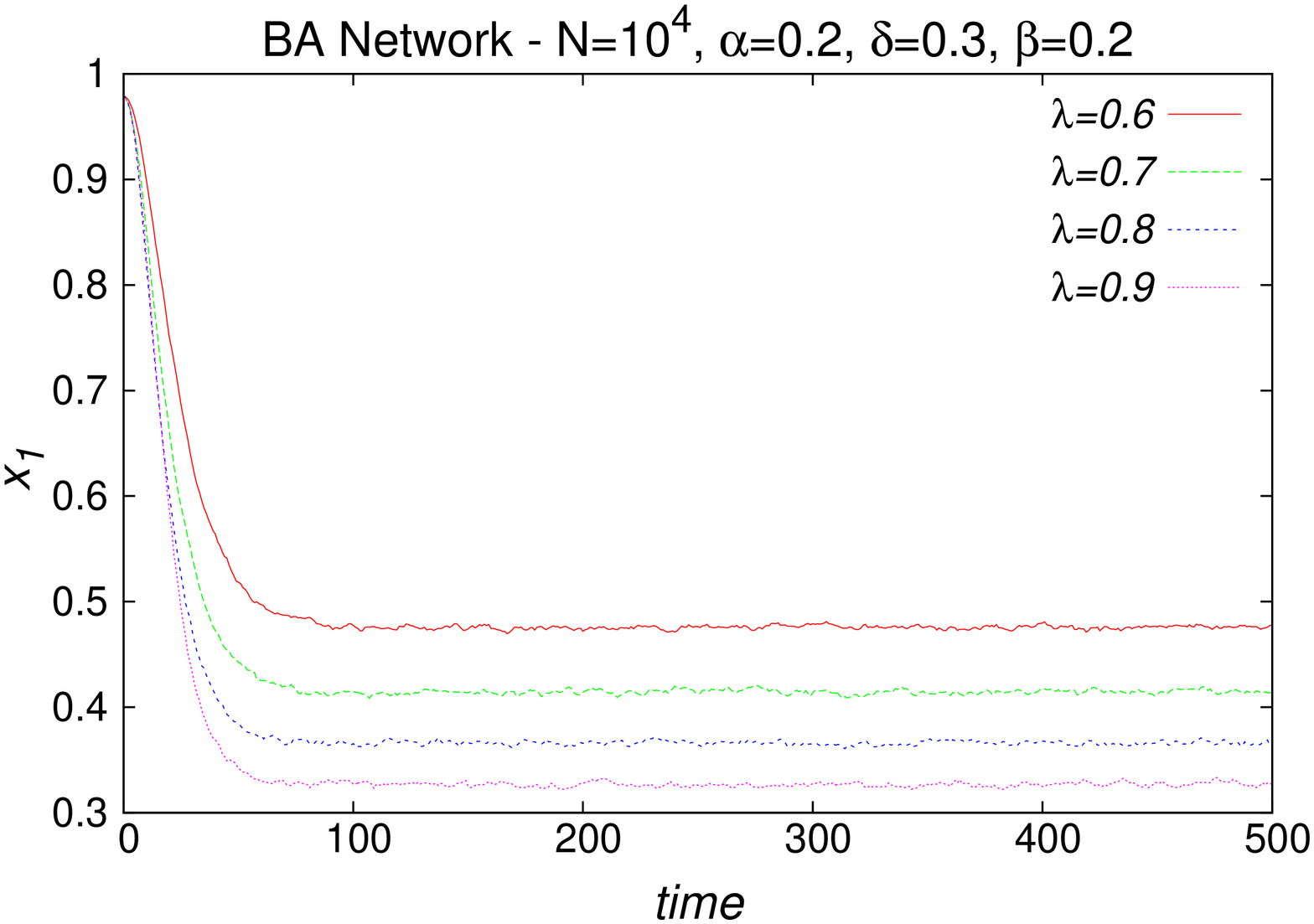}
\hspace{0.3cm}
\includegraphics[width=0.33\textwidth,angle=270]{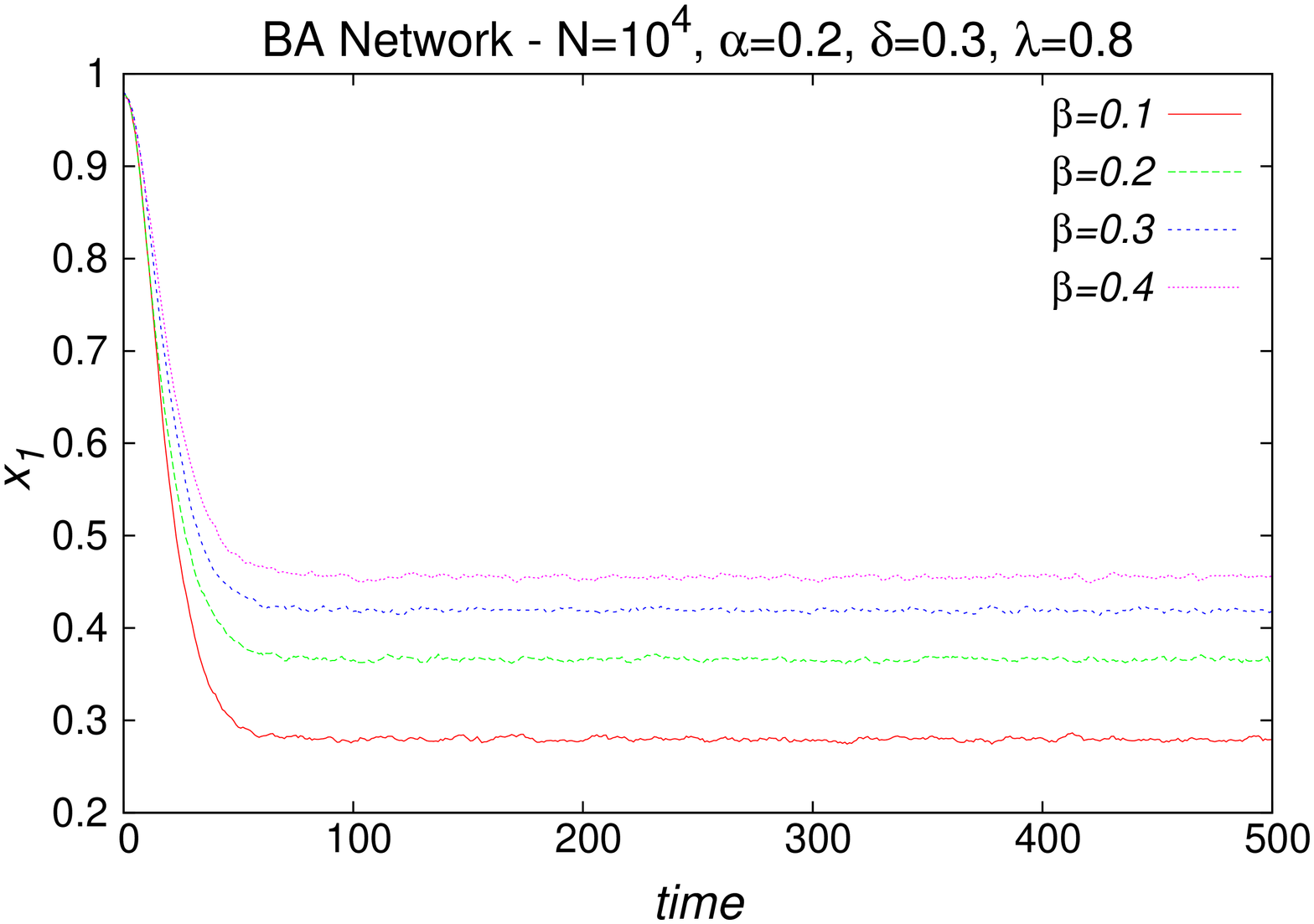}
\\
\vspace{0.5cm}
\includegraphics[width=0.33\textwidth,angle=270]{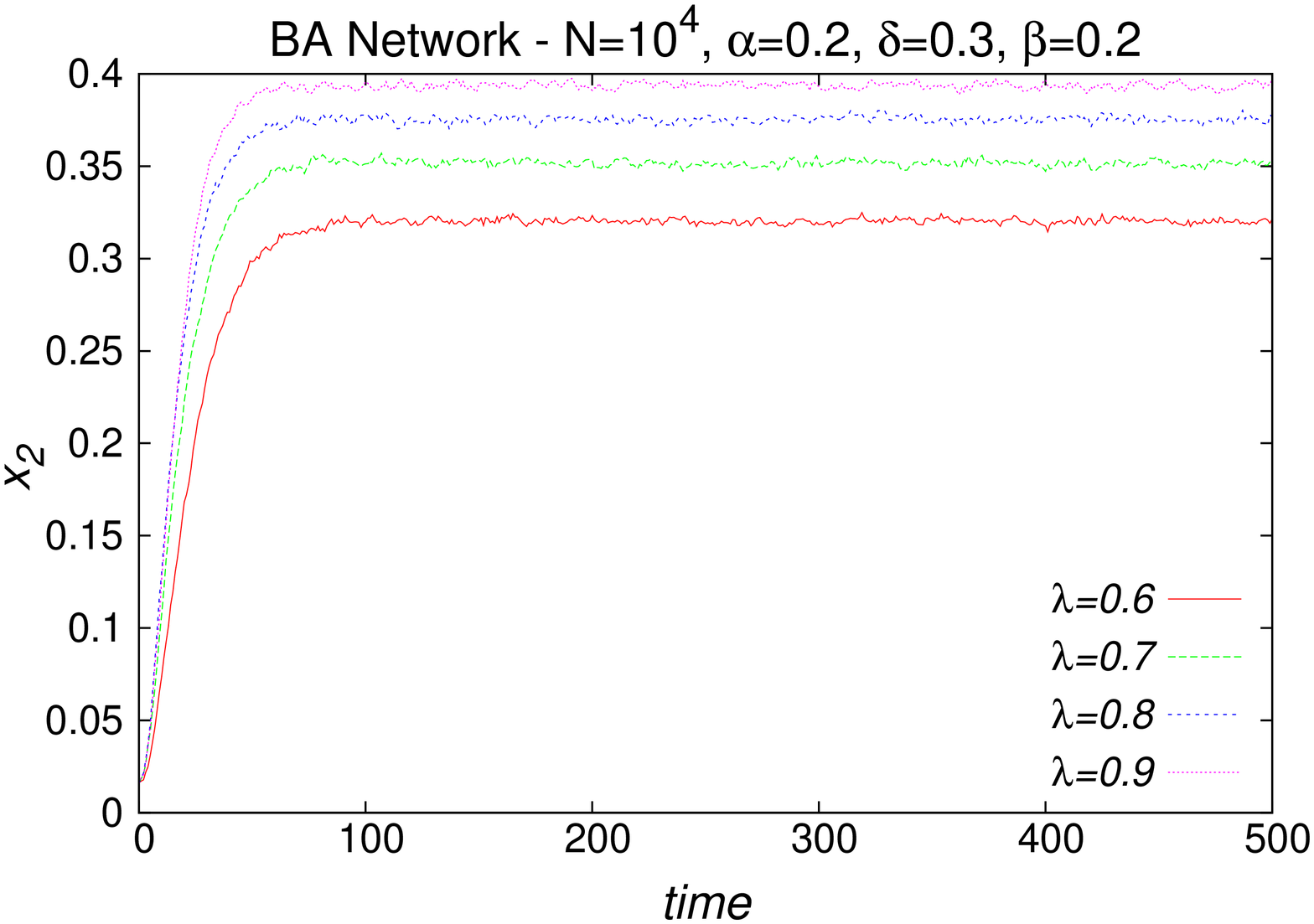}
\hspace{0.3cm}
\includegraphics[width=0.33\textwidth,angle=270]{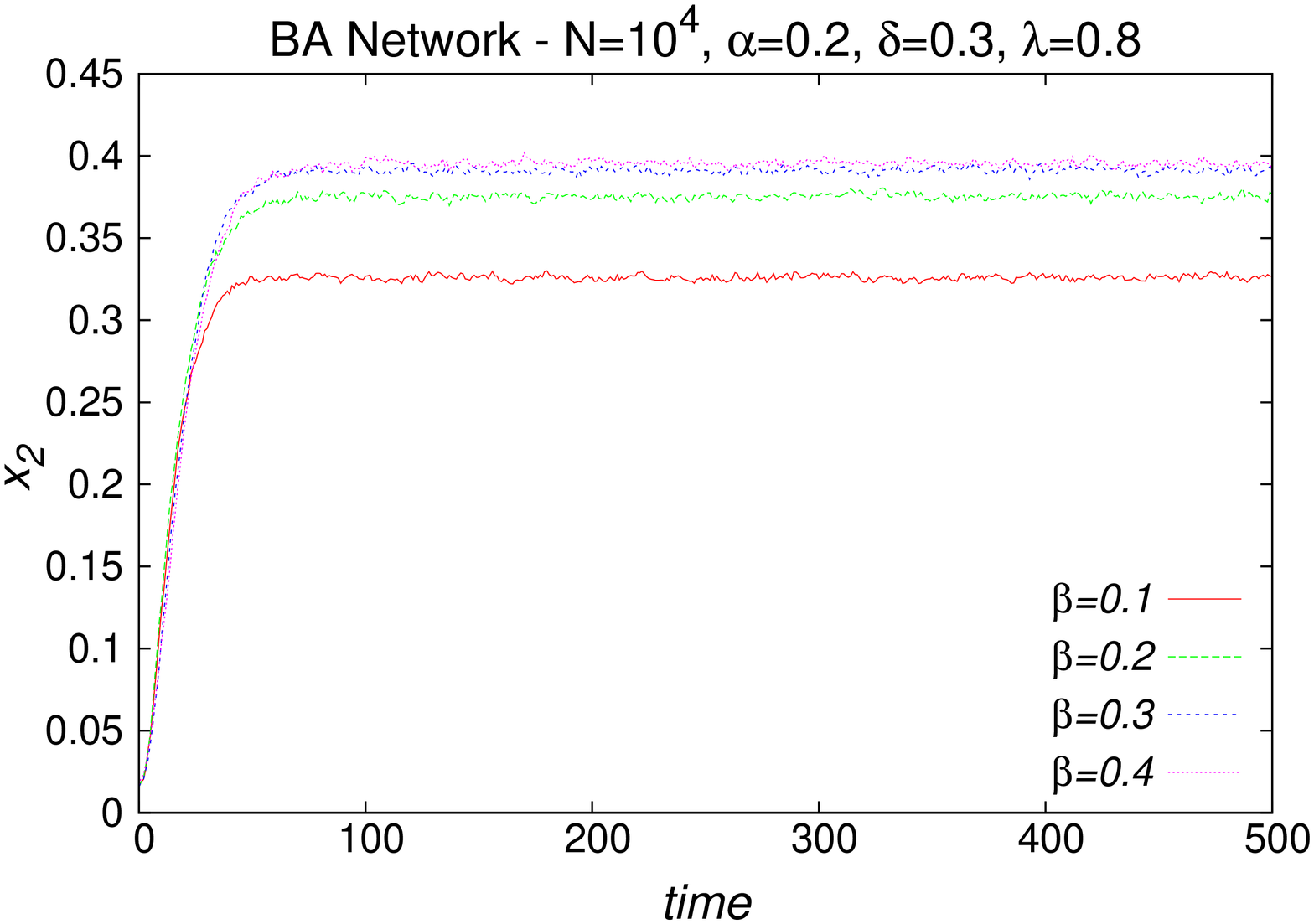}
\\
\vspace{0.5cm}
\includegraphics[width=0.33\textwidth,angle=270]{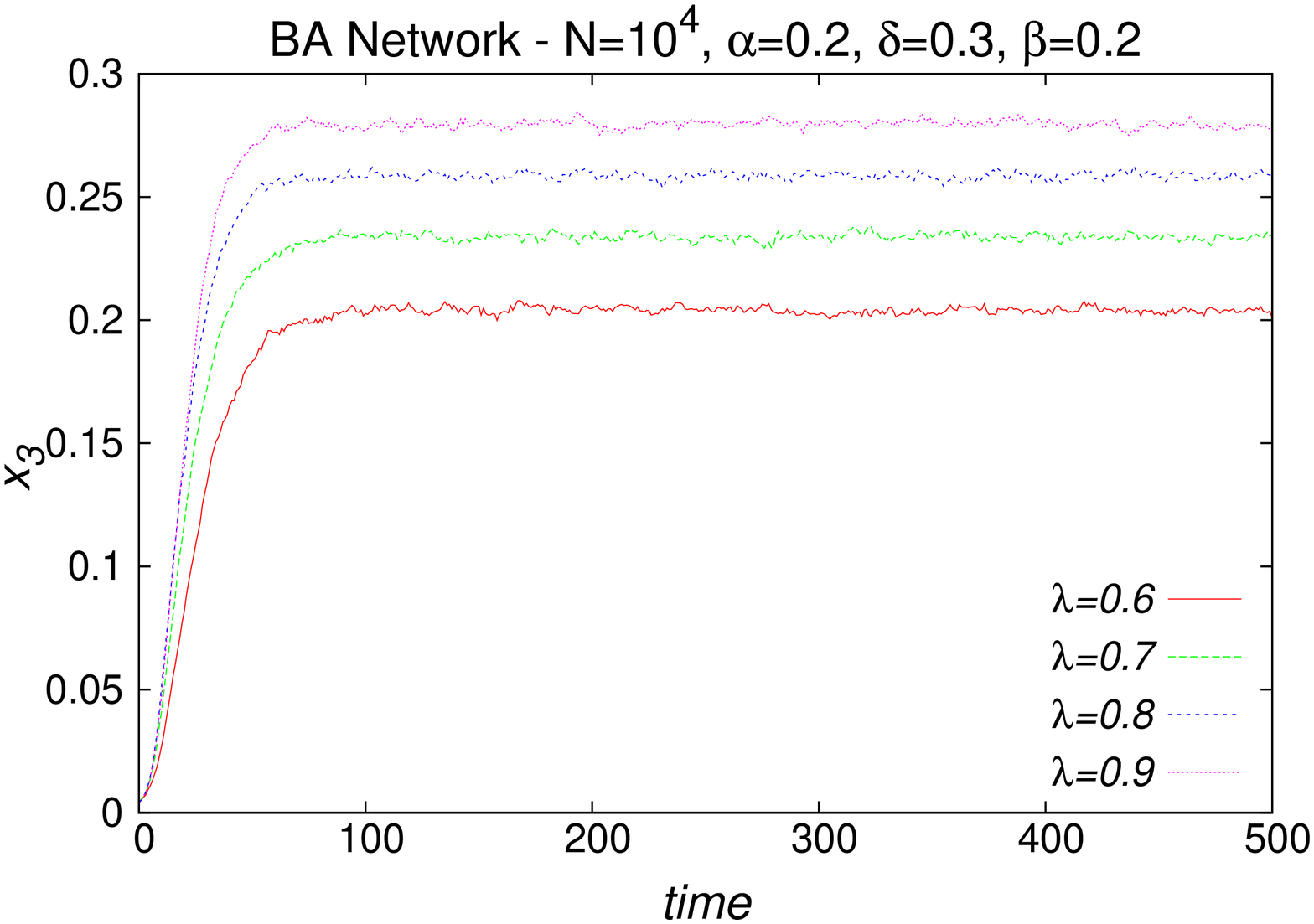}
\hspace{0.3cm}
\includegraphics[width=0.33\textwidth,angle=270]{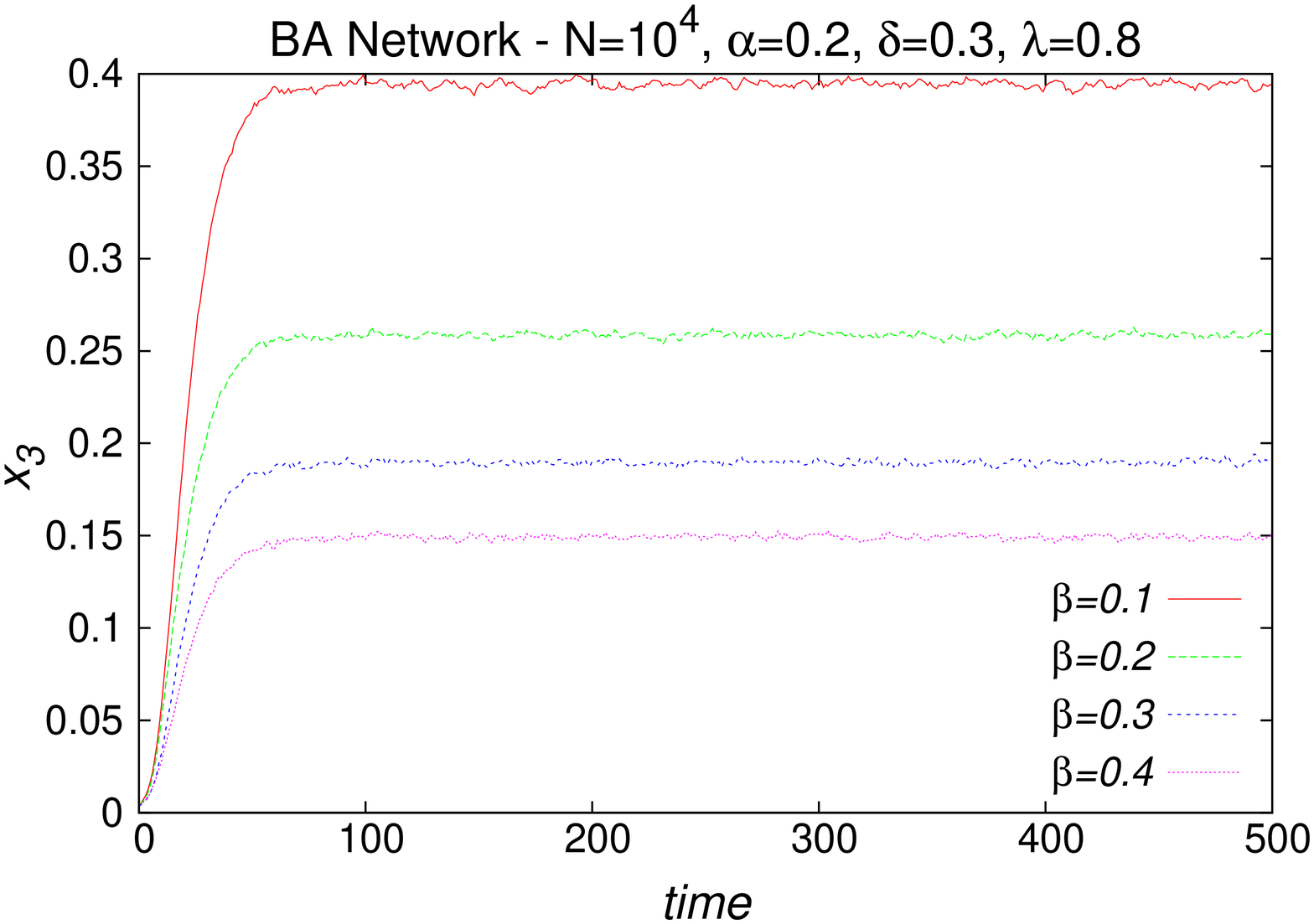}
\end{center}
\caption{(Color online) Time evolution of the three densities of agents $x_{1}$, $x_{2}$ and $x_{3}$ obtained from simulations of the model defined on the BA scale-free networks. The fixed parameters are $\alpha=0.2$ and $\delta=0.3$. In the left panels it is shown the evolution for $\beta=0.2$ and typical values of $\lambda$, whereas in the right panels we exhibit the evolution for $\lambda=0.8$ and typical values of $\beta$.}
\label{fig5}
\end{figure}

\begin{figure}[t]
\begin{center}
\vspace{6mm}
\includegraphics[width=0.33\textwidth,angle=270]{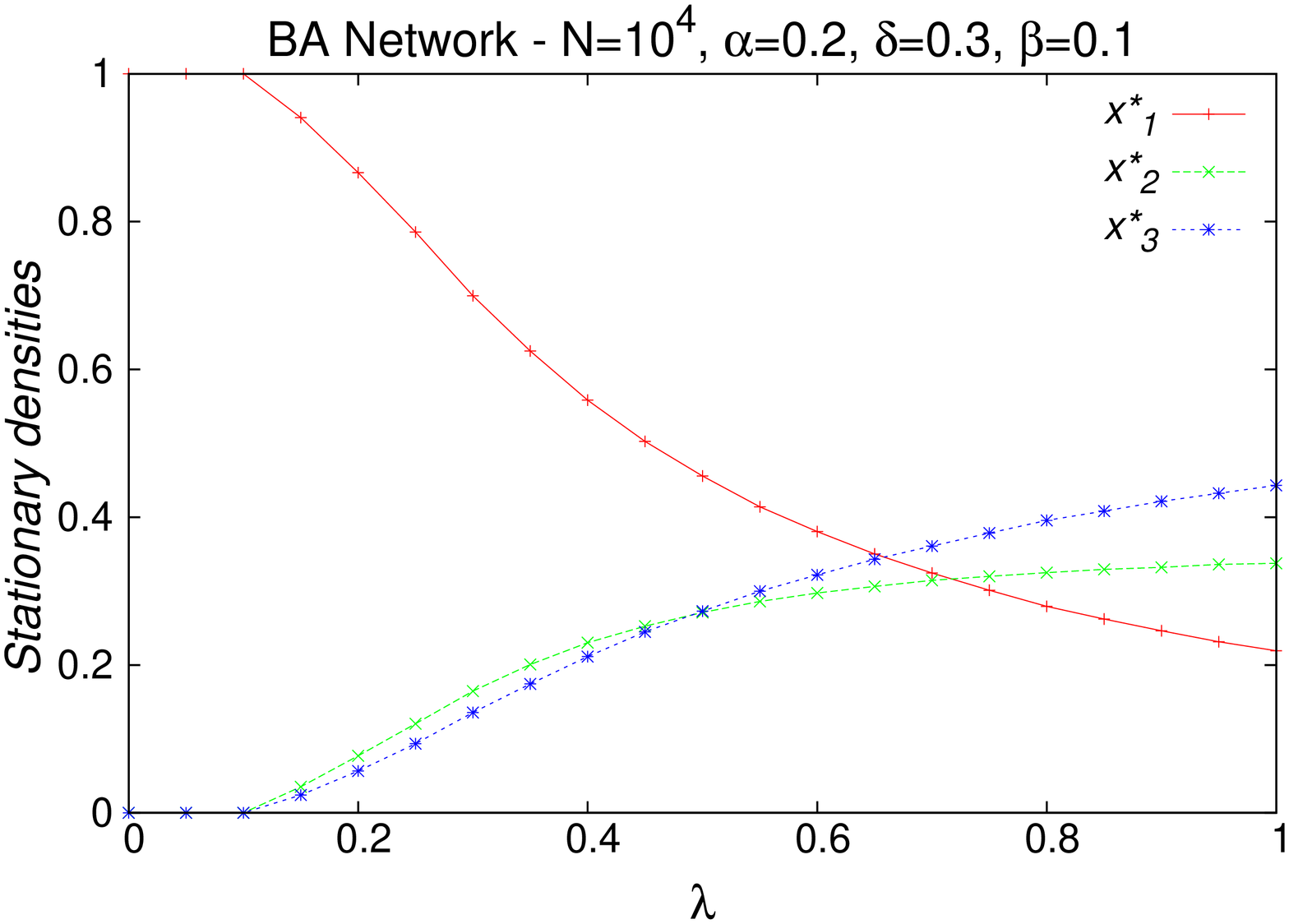}
\hspace{0.3cm}
\includegraphics[width=0.33\textwidth,angle=270]{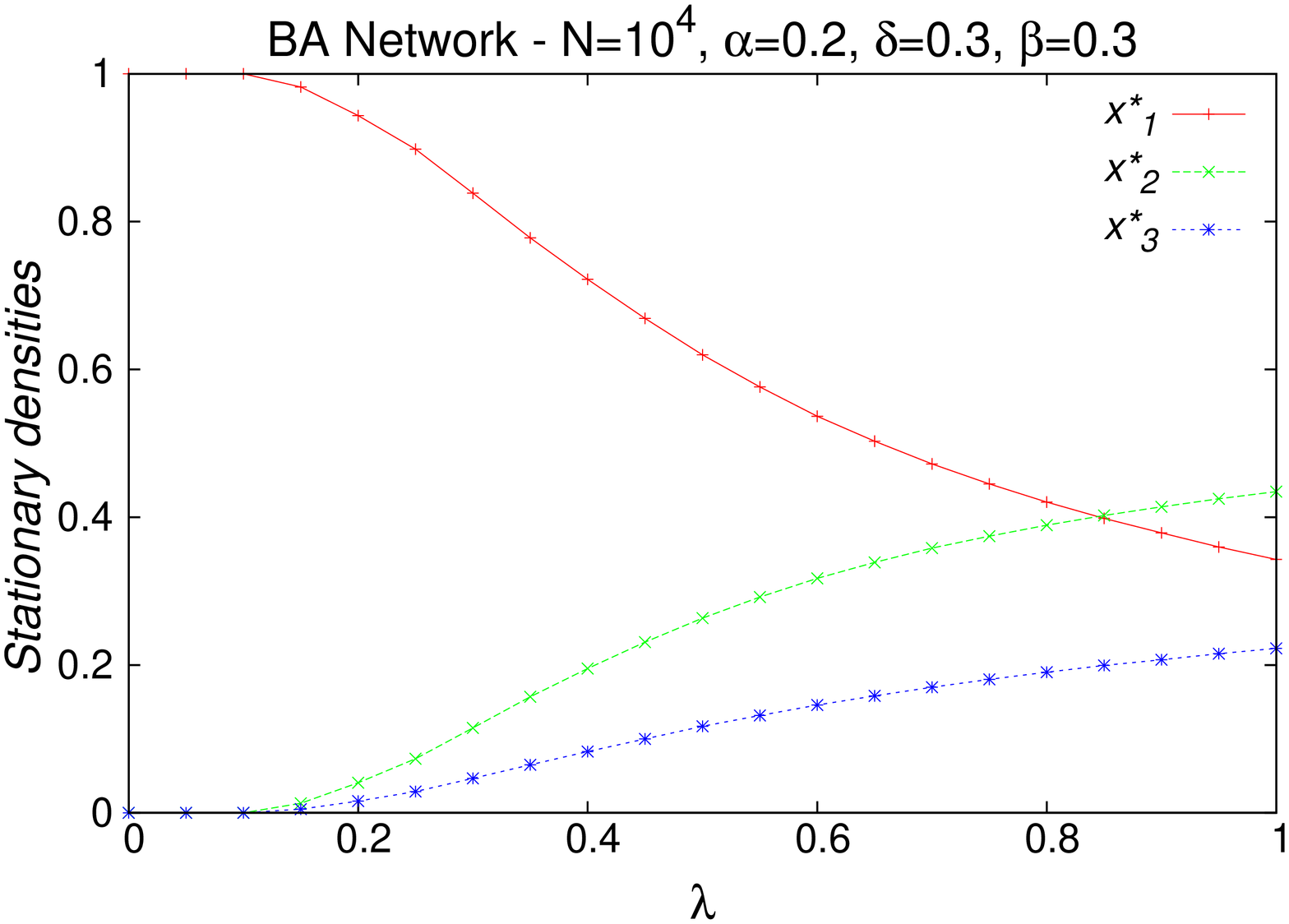}
\\
\vspace{0.5cm}
\includegraphics[width=0.33\textwidth,angle=270]{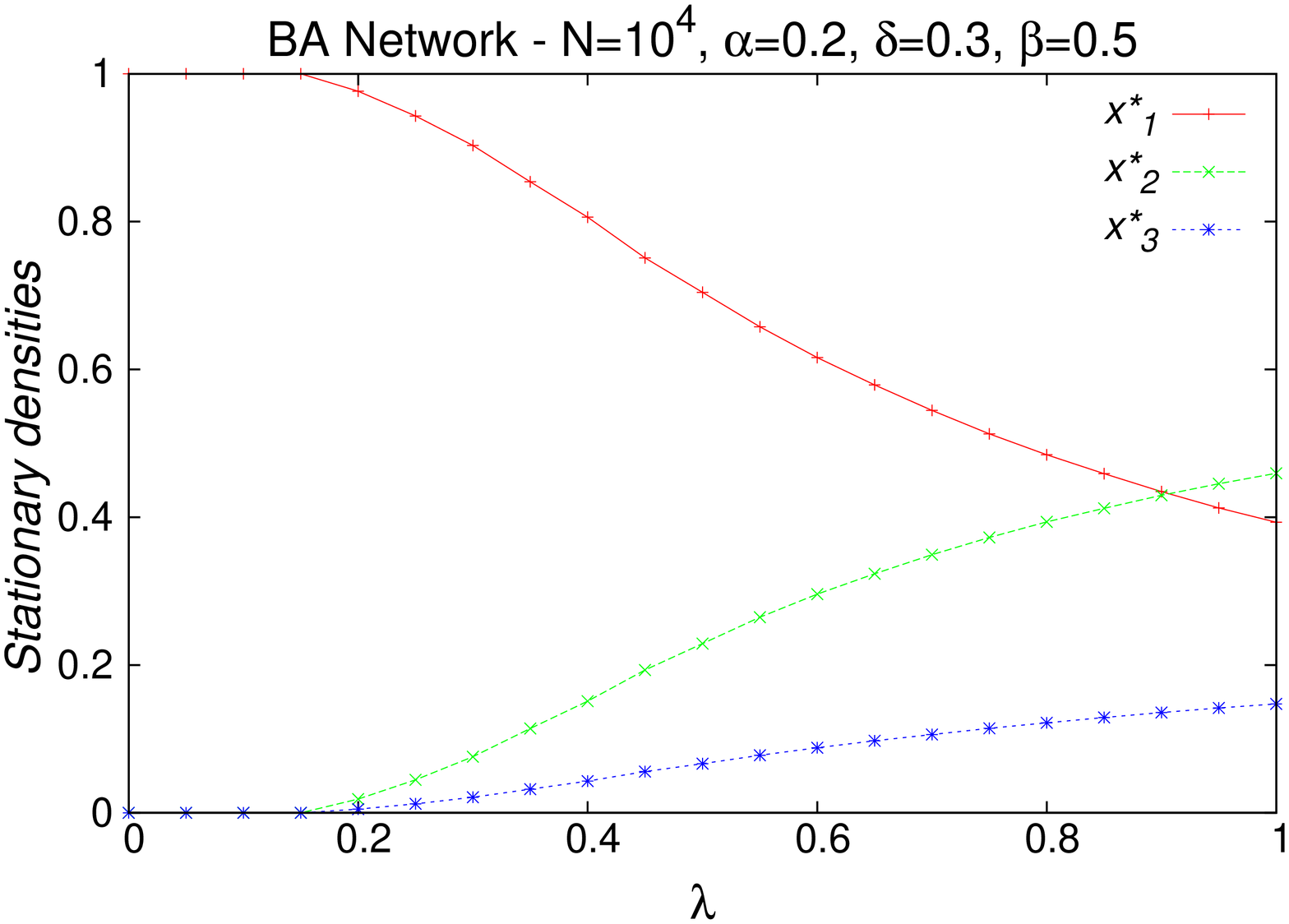}
\end{center}
\caption{(Color online) Stationary fractions $x_{1}^{*}$, $x_{2}^{*}$ and $x_{3}^{*}$ as functions of $\lambda$ for $\beta=0.1$ (upper left), $\beta=0.3$ (upper right) and $\beta=0.5$ (lower) for the model simulated on the BA network. The fixed parameters are $\alpha=0.2$ and $\delta=0.3$.}
\label{fig6}
\end{figure}

The time evolution of the densities is very similar to the previous cases, as it is shown in Fig. \ref{fig5}. In addition, the stationary values $x_{1}^{*}$, $x_{2}^{*}$ and $x_{3}^{*}$ are exhibited in Fig. \ref{fig6}. In comparison with the previous cases, the decrease of the fraction of honests is slower when we rise the probability $\lambda$. We also see a similar behavior related to the increase of the enforcement $\beta$, with a good reduction of tax evasion, and a rapid increase of the susceptible agents for increasing values of $\lambda$.

\begin{figure}[t]
\begin{center}
\vspace{6mm}
\includegraphics[width=0.33\textwidth,angle=270]{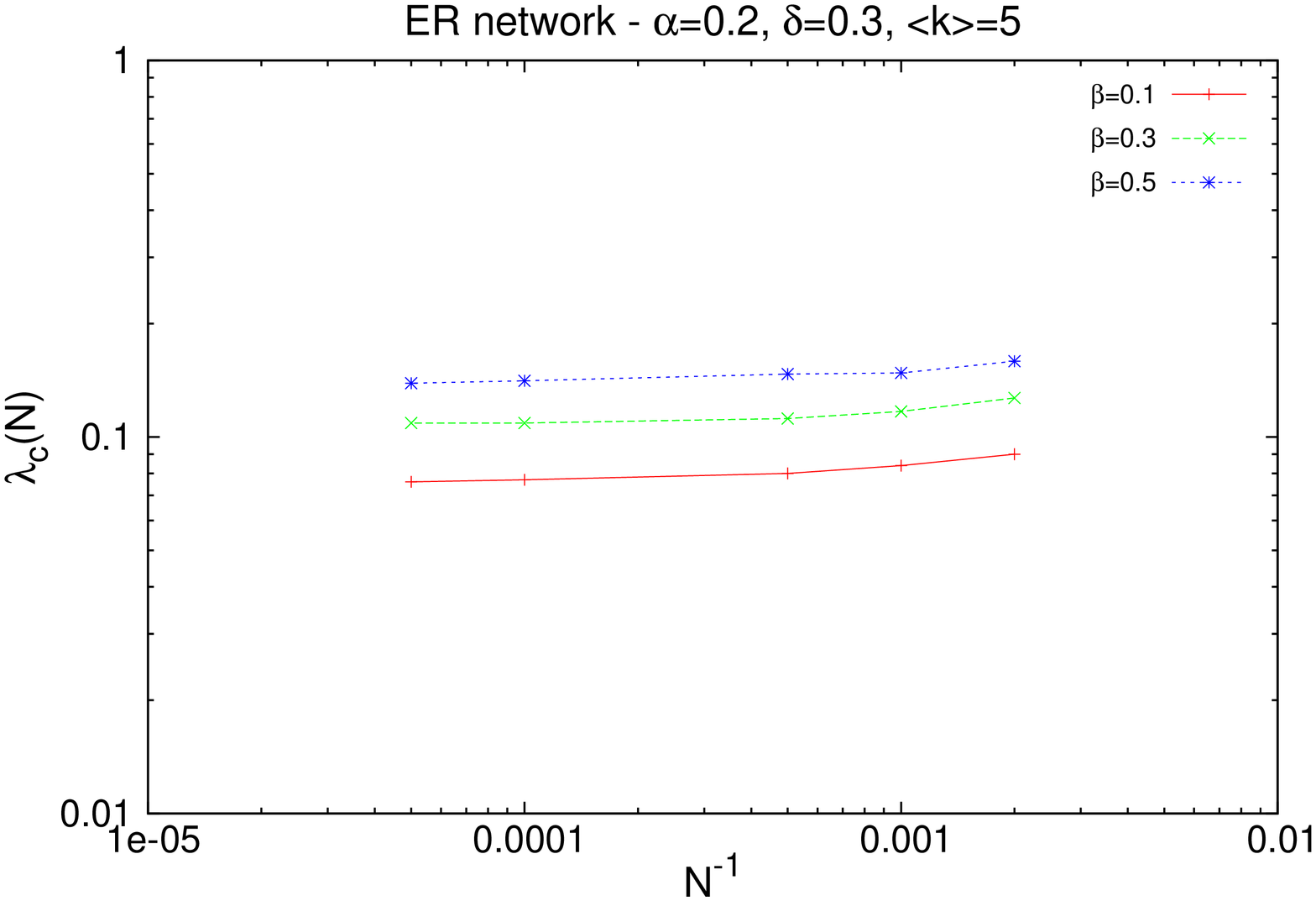}
\hspace{0.3cm}
\includegraphics[width=0.33\textwidth,angle=270]{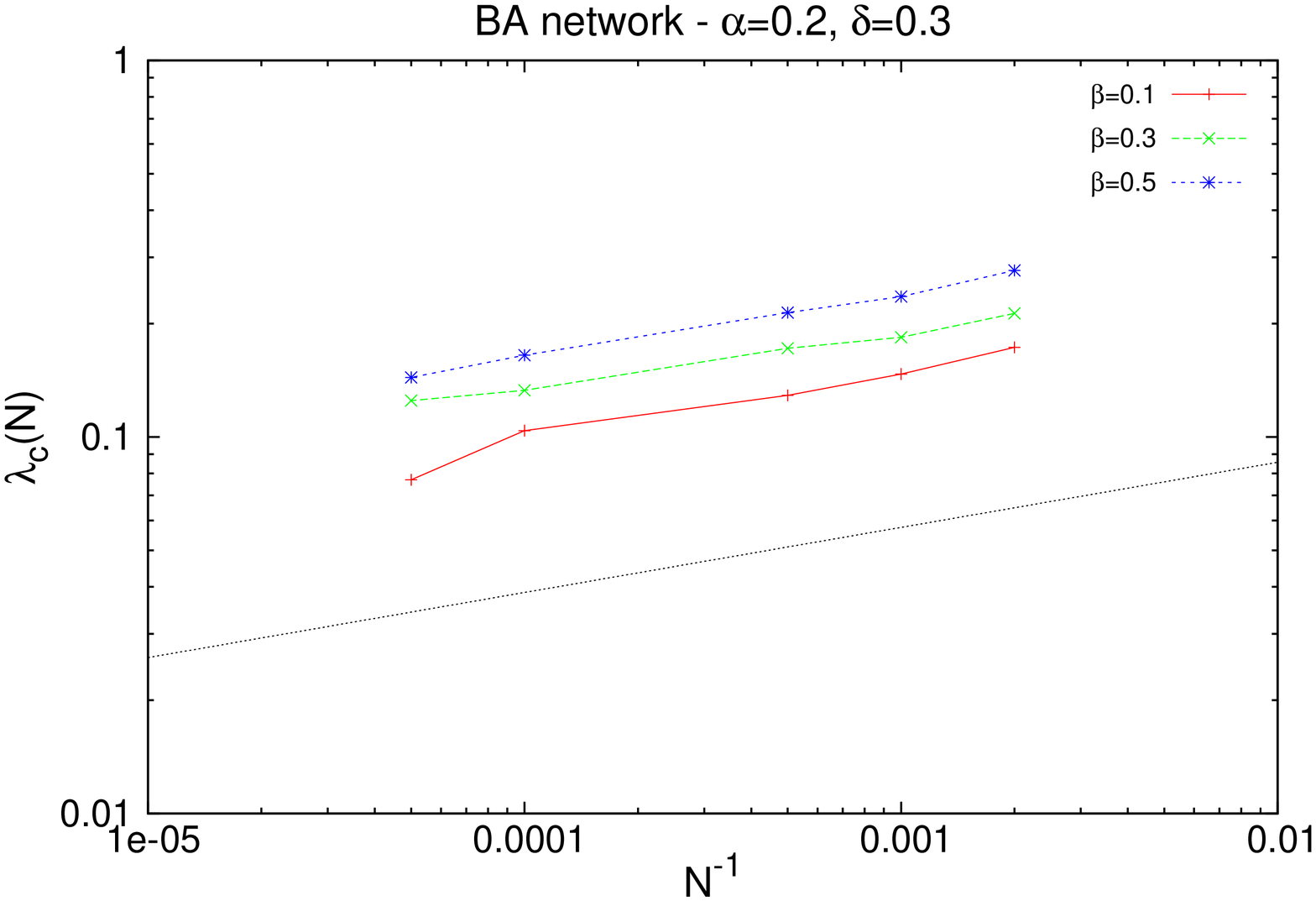}
\end{center}
\caption{(Color online) Thresholds $\lambda_{c}(N)$ as functions of the inverse network size $N^{-1}$ for the ER random graph (left) and BA network (right). The fixed parameters are $\alpha=0.2$ and $\delta=0.3$, and the results are for $\beta=0.1$, $0.3$ and $0.5$. The straight line in panel (b) indicates a power-law decay on $\lambda_{c}(N)$, signaling the absence of the ``epidemic threshold'' in the thermodynamic limit.}
\label{fig7}
\end{figure}

As in the other graphs, one can see that we have $x_{3}^{*}=0$ for sufficient small values of $\lambda$. However, as it is typical in epidemic-like models in scale-free BA networks, this is only a finite size effect, and we do not expect an ``epidemic threshold'' in this case \cite{satorras2015}. As an evidence of this fact, we plot in Fig. \ref{fig7} the thresholds $\lambda_{c}(N)$ obtained from simulations as functions of the inverse networks size $N^{-1}$. As one can see in the left panel of Fig. \ref{fig7}, for the ER random graph $\lambda_{c}(N)$ tends to stabilize at a finite value for increasing values of $N$, suggesting a nonzero threshold. On the other hand, for the BA network the threshold $\lambda_{c}(N)$ decays as a power law for increasing sizes, signaling the absence of the ``epidemic threshold'' in the thermodynamic limit $N^{-1}\to 0$ (see the right panel of Fig. \ref{fig7}) \cite{satorras2015}. In this case, for a sufficient large network, one expect the presence of noncompliant individuals (tax evaders) in the population in the long-time limit, for any value of $\lambda>0$. 

Comparing the results for the two complex topologies (ER and BA), one can see  that a given variation of $\lambda$ (social parameter) leads to distinct variations of the fraction of tax evaders (see Figs. \ref{fig3} - \ref{fig6}). This effect can be related to the presence of hubs (highly connected nodes) in the BA network, as well as the large fluctuation in the average connectivity. These characteristics are absent in the ER random graph, and are crucial for the spreading of influence in the population.

\section{Concluding remarks}   

\qquad In this work, we have studied the dynamics of tax evasion through an epidemic-like model. We considered three compartments, namely honest tax payers, tax evaders and susceptibles, individuals that are in a intermediate class among honests and dishonests. The transitions among these classes are ruled by four distinct probabilities, that represent social interactions and the enforcement regime. We study the dynamics of the system on the top of three distinct topologies: fully-connected graph, Erd\"{o}s-R\'enyi random graph and Barab\'asi-Albert scale-free network.

For the fully-connected graph, one derive mean-field equations that allow us to analyze in details the dynamics and the steady-state properties of the model. Some realistic behaviors were observed, as the reduction of the evasion due to enforcement regime, as well as the increse of honesty. We also observed that the emergence of evaders in the population is associated with a nonequilibrium phase transition: for small values of the control parameter there are only honest tax payers in the population, and above the critical point the three classes (honests, susceptibles and evaders) coexist in the system. The results are qualitatively similar for the other topologies, but regarding the stationary values of the densities of individuals, we verified that the control of tax evasion is harder if the model is simulated on the top of complex networks. In this case, we also verified that the effect of social pressure is more pronounced in comparison with the mean-field case. 

We observed that in the mean-field case the tax evasion (fraction of noncompliant agents) is absent ($x_{3}^{*}=0$) even for high social pressure (high $\lambda$), which can be seen as a limitation of the model in the simple case of a fully-connected topology. However, this unrealistic feature disappears when we considered the model on the top of complex networks. Indeed, for the ER random graph the region where the tax evaders disappear of the population in the steady states is given by a narrow range of values of $\lambda$, even if the government's fiscalization is high. On the other hand, in the BA scale-free network this effect is more pronounced, and for sufficient large networks the evasion is always present in the population, even for strong fiscalization.

Some qualitative comparison with real data can be done. Some authors  estimated the tax evasion in Brazil in the range $15 - 22\%$, or even higher values (see \cite{siqueira} and references therein). This range of evasion can be verified in our results for small values of $\beta$, i.e., for weak fiscalization and/or light punishment, as occurs in Brazil \cite{page1,book}. For example, in Fig. \ref{fig2} the fraction of evasion in the range $15 - 22\%$ can be observed for $\beta=0.1$, considering the range $\approx 0.45 < \lambda < 0.5$ (mean field). For the networks, one can see the mentioned range of evasion for $\approx 0.15 < \lambda < 0.4$ ($\beta=0.1$) and $\approx 0.2 < \lambda < 0.7$ ($\beta=0.3$), for the ER network (see Fig. \ref{fig4}). For the BA network, $\approx 0.25 < \lambda < 0.4$ ($\beta=0.1$) and $\approx 0.5 < \lambda < 1.0$ ($\beta=0.3$) (see Fig. \ref{fig6}).

The phase transition observed in the mean-field formulation of the model is an active-absorbing phase transition, and the predicted exponent for the order parameter is $1$ ($x_{3}^{*} \sim (\lambda-\lambda_{c})^{1}$) as in the mean-field directed percolation, that is the prototype of a phase transition to an absorbing state \cite{dickman,hinrichsen}. It would be interesting to estimate numerically other critical exponents of the model, as well as to simulate it in regular d-dimensional lattices (square, cubic, for example) in order to obtain all the critical exponents. This is important to define precisely the universality class of the model, as well as its upper critical dimension. This extension is left for a future work. Furthermore, it can also be considered the inclusion of heterogeneities in the population, like agents' conviction, mass media effects, etc.


\section*{Acknowledgments}

The authors acknowledge financial support from the Brazilian funding agencies CNPq and CAPES.

\end{document}